\documentclass[twocolumn]{aastex6}
\hypersetup{linkcolor={blue},citecolor={blue}}
\shorttitle{SN EJECTA WITH A RELATIVISTIC WIND}
\shortauthors{SUZUKI\&MAEDA}
\begin{document}
\title{SUPERNOVA EJECTA WITH A RELATIVISTIC WIND FROM A CENTRAL COMPACT OBJECT: A UNIFIED PICTURE FOR EXTRAORDINARY SUPERNOVAE}
\author{AKIHIRO SUZUKI\altaffilmark{1} and KEIICHI MAEDA\altaffilmark{1,2}}
\altaffiltext{1}{Department of Astronomy, Kyoto University, Kitashirakawa-Oiwake-cho, Sakyo-ku, Kyoto, 606-8502, Japan}
\altaffiltext{2}{Kavli Institute for the Physics and Mathematics of the Universe (WPI),
University of Tokyo, 5-1-5 Kashiwanoha, Kashiwa, Chiba 277-8583, Japan}
\begin{abstract}
The hydrodynamical interaction between freely expanding supernova ejecta and a relativistic wind injected from the central region is studied in analytic and numerical ways. 
As a result of the collision between the ejecta and the wind, a geometrically thin shell surrounding a hot bubble forms and expands in the ejecta. 
We use a self-similar solution to describe the early dynamical evolution of the shell and carry out a two-dimensional special relativistic hydrodynamic simulation to follow further evolution. 
The Rayleigh-Taylor instability inevitably develops at the contact surface separating the shocked wind and ejecta, leading to the complete destruction of the shell and the leakage of hot gas from the hot bubble. 
The leaking hot materials immediately catch up with the outermost layer of the supernova ejecta and thus different layers of the ejecta are mixed. 
We present the spatial profiles of hydrodynamical variables and the kinetic energy distributions of the ejecta. 
We stop the energy injection when a total energy of $10^{52}$ erg, which is 10 times larger than the initial kinetic energy of the supernova ejecta, is deposited into the ejecta and follow the subsequent evolution. 
From the results of our simulations, we consider expected emission from supernova ejecta powered by the energy injection at the centre and discuss the possibility that superluminous supernovae and broad-lined Ic supernovae could be produced by similar mechanisms. 
\end{abstract}

\section{INTRODUCTION\label{intro}}
A recently found special class of supernovae (SNe) producing 10-100 times brighter emission than normal supernovae is called superluminous supernovae (SLSNe) and poses a big theoretical challenge in explaining their origin \citep[see,][for review]{2012Sci...337..927G}. 
SLSNe lacking any hydrogen feature in their spectra are classified as type-I SLSNe (SLSNe-I) in an analogy to the classification scheme for normal SNe \citep{2007ApJ...668L..99Q,2009ApJ...690.1358B,2010ApJ...724L..16P,2011Natur.474..487Q,2011ApJ...743..114C}. 
Their spectral features suggest that they probably originate from massive stars having lost their hydrogen- and helium-rich layers. 
However, the mechanism to produce the tremendous amount of radiation observed for these enigmatic events is still unclear. 

Among their characteristic features, the rarity and the preference to extreme environments may hint at their origin. 
The volumetric rate of SLSNe-I has independently been measured by several groups. 
From the SLSN samples found by Robotic Optical Transient Search Experiment-IIIb (ROTSE-IIIb) telescope observations, 
\cite{2013MNRAS.431..912Q} measured the SLSN-I volumetric rate at $z\simeq 0.17$ to be $32^{+77}_{-26}\ \mathrm{yr}^{-1}\ \mathrm{Gpc}^{-3}$. 
In the samples of Supernova Legacy Survey \citep{2006A&A...447...31A} found in 2003-2008, \cite{2016arXiv160505250P} identified three SLSNe around $z\sim 1$, including two samples having been reported previously \citep{2013ApJ...779...98H}. 
They calculated the volumetric rate of SLSNe-I at $z\sim 1$ to be $91^{+76}_{-36}\ \mathrm{yr}^{-1}\ \mathrm{Gpc}^{-3}$. 
These rates are significantly smaller than those of normal stripped-envelope core-collapse SNe at the corresponding redshifts. 
Pan-STARRS1 observations of SLSNe-I \citep{2015MNRAS.448.1206M} did not measure the volumetric rate, but they estimated the relative rate to normal core-collapse SNe to be between $3^{+3}_{-2}\times10^{-5}$ and $8^{+2}_{-1}\times10^{-5}$. 

Recent observations of host galaxies of SLSNe-I revealed that they tend to occur in dwarf galaxies with high specific star formation rates and low metallicity \citep{2011ApJ...727...15N,2014ApJ...787..138L,2015MNRAS.449..917L,2016arXiv160504925C}. 
The first systematic study of SLSNe-I host galaxies by \cite{2014ApJ...787..138L} indicates that they have a lot in common with galaxies hosting long-duration gamma-ray bursts (GRBs). 
\cite{2015MNRAS.449..917L} reported that a considerable fraction of SLSNe-I host galaxies is classified as Extreme Emission Line Galaxies, which exhibit strong emission lines, e.g., [\ion{O}{3}], in their spectra \citep{2011ApJ...743..121A}. 
They argued that the progenitor system producing SLSNe-I must be closely linked with star-forming activity in metal-poor environments.

In addition to their still enigmatic origin, the high brightness of SLSNe makes them even more attractive for astronomers because they could be detected at the high-$z$ universe. 
Simulations of the detectability of SLSNe at the high-$z$ universe \citep{2012MNRAS.422.2675T,2013MNRAS.435.2483T} predicted that currently ongoing and upcoming surveys, such as Subaru/Hyper Suprime-Cam \citep{2012SPIE.8446E..0ZM}, Euclid, The Wide-Field Infrared Survey Telescope (WFIRST), and Wide-field Imaging Surveyor for High-redshift (WISH), would further increase the number of SLSN detections. 
\cite{2012Natur.491..228C} actually discovered two possible SLSNe at redshifts $z= 2.1$ and $3.9$. 
\cite{2014ApJ...796...87I} analysed their SLSNe-I samples and claimed that SLSNe-I can be used as a distant indicator at the high-$z$ universe. 
The preference for extreme environments and the high brightness of SLSNe strongly indicate their potential to probing star-forming activity at the high-$z$ universe. 

From theoretical points of view, several scenarios for possible energy sources of SLSNe-I have been proposed. 
Widely-discussed scenarios are the radioactive decay of $^{56}$Ni produced in pair-instability supernovae (PISNe), the interaction between supernova ejecta and dense hydrogen deficit circum-stellar media (CSM), and an additional energy injection from a compact remnant. 
First, massive stars with initial masses of $\sim(140-300)\ M_\odot$ can be dynamically unstable by creating electron-positron pairs at the core, leading to the complete disruption of the star \citep{1967PhRvL..18..379B,1967ApJ...148..803R,2002ApJ...567..532H}. 
Some luminous events have been claimed to be powered by radioactive decay of $^{56}$Ni abundantly produced in the explosion \citep[e.g.][]{2009Natur.462..624G}. 
Next, in the CSM interaction models, collisions between supernova ejecta and a dense CSM or a shell having been ejected from the progenitor star lead to efficient shock heating, giving rise to bright emission. 
The dense material surrounding the exploding star can be provided by a stellar wind at a high mass-loss rate \citep{2011ApJ...729L...6C} or pulsational pair-instability prior to the explosion \citep{2007Natur.450..390W,2012ApJ...760..154C,2016MNRAS.457..351Y}.  
Finally, the compact object left in the supernova ejecta can also supply the expanding ejecta with an additional energy. 
The energy injection may be realized by a new-born rapidly rotating magnetized neutron star through its spin-down (\citealt{2010ApJ...717..245K,2010ApJ...719L.204W}; see also \citealt{1971ApJ...164L..95O,1976SvA....19..554S,2007ApJ...666.1069M}) or accretion onto a stellar mass black hole \citep{2013ApJ...772...30D}. 

Light curve and spectral modellings have been extensively carried out for these scenarios, PISNe \citep{2011ApJ...734..102K,2012MNRAS.426L..76D,2013MNRAS.428.3227D,2014A&A...565A..70K,2014A&A...566A.146K,2015MNRAS.454.4357K}, CSM interaction \citep{2007Natur.450..390W,2007ApJ...671L..17S,2011ApJ...729L...6C,2012ApJ...757..178G,2013MNRAS.428.1020M,2015MNRAS.449.4304D,2016ApJ...829...17S}, and magnetar energy injection \citep{2010ApJ...717..245K,2012MNRAS.426L..76D}. 
One-zone analytic light curve models integrating these three energy reservoirs have been formulated \citep[e.g.][]{2012ApJ...746..121C} and applied to observed light curves of SLSNe-I in comprehensive ways \citep{2013ApJ...770..128I,2013ApJ...773...76C,2015MNRAS.452.3869N}. 
These studies revealed that at least some SLSNe-I are unlikely to be powered by $^{56}$Ni due to their relatively short durations, high brightness, and absence of exponential tails in their light curves. 
In addition, the discovery of the brightest SLSN, ASASSN-15lh \citep{2016Sci...351..257D}, has invoked active debates on its energy reservoir \citep{2015MNRAS.454.3311M,2016ApJ...817L...8B,2016ApJ...817..132D,2016ApJ...820L..38S,2016MNRAS.459L..21K}, although this event is suspected to be a tidal disruption event \citep[see discussions in][]{2016arXiv160500645G,2016arXiv160503951B,2016arXiv160902927L,2016arXiv161001632M}

Another key issue is the connection between SLSNe-I and ultra-long GRBs. 
A population of GRBs with exceptionally long duration, $T_{90}\simeq 10^3$-$10^4$ s, has been identified and is termed ultra-long GRBs \citep[see, e.g.,][]{2014ApJ...781...13L}. 
Recently, \cite{2015Natur.523..189G} reported that a supernova-like transient, which was named SN 2012kl, was associated with the afterglow of the ultra-long GRB 111209A. 
SN 2012kl was found to be at least three times more luminous than other energetic SNe associated with GRBs. 
The high peak luminosity and relatively short duration, several 10 days, are difficult to explain in the standard $^{56}$Ni powered emission model, making this event another candidate for SNe with an additional energy supply. 

Therefore, investigations on supernova ejecta with a central engine are of crucial importance in revealing the origin of SLSNe-I and the mechanism responsible for their bright emission. 
The dynamical evolution of expanding supernova ejecta with an additional energy injection at the centre has been considered for a long time, in the context of the evolution of a pulsar wind nebula embedded in an SN remnant (e.g., \citealt{1984ApJ...278..630R,1984ApJ...280..797C,1984ApJ...283..694K,1998ApJ...499..282J,2001A&A...380..309V,2001ApJ...563..806B}; see \citealt{2006ARA&A..44...17G} for review). 
In these studies, the total amount of the deposited energy is usually less than the explosion energy of the SN. 
On the other hand, for SLSNe-I, the energy injected from the central engine should eventually overwhelm the kinetic energy of the supernova ejecta because the total radiated energy of SLSNe often reaches $10^{51}$ erg. 
Although several studies in this regime have been carried out \citep[e.g.][]{2010ApJ...717..245K}, most of them assume spherical symmetry and thus some multi-dimensional effects could be overlooked. 
One-dimensional calculations show that the energy injection at the bottom of supernova ejecta leads to the formation of a geometrically thin shell in the ejecta. 
However, multi-dimensional hydrodynamic simulations of supernova remnants with pulsar wind nebulae \citep{1998ApJ...499..282J,2001ApJ...563..806B} have revealed that the Rayleigh-Taylor instability develops around the interface between the nebula and the supernova ejecta, which modifies the internal structure of the ejecta. 
In addition, a thin shell surrounded by a couple of shock fronts is known to be subjected to a variety of hydrodynamic instabilities, such as Richtmyer-Meshkov instability \citep{R60,1972FlDy....4..101M} and non-linear thin shell instability \citep{1994ApJ...428..186V}. 
Furthermore, the energy injection may be realized in aspherical ways. 
In the context of GRBs, some studies have considered central engine activities after supernova ejecta have been created \citep[e.g.][]{1998ApJ...507L..45V}. 
In such situation, the energy injection is realized by the launch of a highly collimated jet or an outflow. 
\cite{2011MNRAS.411.2054L} analytically investigated the launch of a bipolar outflow and the propagation of the blast wave in supernova ejecta.

Particularly, in the magnetar spin-down scenario, the gas injected from the magnetized neutron star into the supernova ejecta is likely to be relativistic as in the case of pulsar wind nebulae. 
Rapidly rotating and highly magnetized neutron stars are also possible acceleration sites of ultrahigh energy cosmic-rays and therefore have received a lot of attention (\citealt{2000ApJ...533L.123B}; \citealt{2003ApJ...589..871A}; \citealt{2009PhRvD..79j3001M}; \citealt{2012ApJ...750..118F,2013JCAP...03..010F}; \citealt{2015JCAP...08..026K}; see \citealt{2011ARA&A..49..119K} for review). 
The hydrodynamic interaction between the wind and the surrounding supernova ejecta has a significant influence on how highly energetic photons, electrons, positrons, and ions likely produced in the magnetosphere of the neutron star escape into the interstellar space. 
\cite{2003ApJ...589..871A} pointed out that the Rayleigh-Taylor instability developing around the interface between the wind and the supernova ejecta could create low-density channels connecting the inner and outer regions in the ejecta, through which high-energy particles can easily escape into the interstellar space. 
Therefore, the dynamical evolution of the supernova ejecta powered by a relativistic wind is of crucial importance in revealing the escape fraction of high-energy particles from the surrounding supernova ejecta. 

Recently, \cite{2016arXiv160407989C} performed 2D non-relativistic hydrodynamic simulations of supernova ejecta with an additional energy injection. 
Their results clearly indicate that the energy injection results in the destruction of the shell and the efficient mixing of layers having been stratified in the ejecta. 
In their simulations, gas injected at the centre of the supernova ejecta travels at non-relativistic speeds. 
However, the energy deposition from the compact object may be realized as an injection of relativistic gas in a similar way to pulsar winds. 
Furthermore, they stopped their calculations shortly after the shell is destroyed and did not follow the further evolution leading to the complete mixing of the ejecta and the injected gas.

In this study, we consider the dynamical evolution of supernova ejecta with an additional energy injection from a central compact object in the form of a relativistic wind. 
We develop a one-dimensional semi-analytic model based on self-similar solutions. 
Furthermore, we perform a numerical simulation by using our 2D special relativistic hydrodynamics code to reveal how the injected energy is distributed throughout the ejecta. 
This paper is organised as follows. 
In Section 2, we describe our assumptions on the supernova ejecta and the relativistic wind. 
In Section 3, we present a semi-analytic model describing the interaction between the wind and the ejecta with spherical symmetry. 
We perform a hydrodynamic simulation to investigate further evolution of the system. 
The setup and results of the numerical simulation are described in Section 4 and 5. 
In Section 6, we discuss the potential of the engine-powered supernova ejecta in producing bright emission and examine the possibility that SLSNe-I and broad-lined Ic SNe originate from supernova ejecta powered by a central engine. 
Finally, we summarise our study in Section 7. 
We describe the derivation of the self-similar solution in Appendix \ref{sec:self_similar} and our numerical code in Appendix \ref{sec:code}. 
We adopt the unit $c=1$, where $c$ is the speed of light unless otherwise noted. 

\section{SUPERNOVA EJECTA AND RELATIVISTIC GAS INJECTION}\label{setup}
\subsection{Ejecta Profile}\label{sec:ejecta_profile}

The supernova ejecta are assumed to be expanding originally in a spherical and homologous way, i.e., the radial velocity is proportional to the radius $R$ (we denote the distance from the origin or the radius in spherical geometry by the capital letter $R$, while the radial coordinate in cylindrical geometry by $r$). 
Thus, the velocity profile at time $t$ is given by
\begin{equation}
v(t,R)=\left\{\begin{array}{ccl}
R/t&\mathrm{for}&R\leq v_\mathrm{ej}t,\\
0&\mathrm{for}&v_\mathrm{ej}t<R,
\end{array}\right.
\label{eq:ejecta_velocity}
\end{equation}
where $v_\mathrm{ej}$ denotes the maximum velocity of the ejecta. 
We adopt the following widely used density profile \citep[e.g.][]{1999ApJS..120..299T}. 
The ejecta are composed of two components, the inner one with a shallow density gradient (referred to as the ``inner ejecta'') and the outer one with a steep density gradient surrounding the inner ejecta (referred to as the ``outer ejecta''). 
The density structures of both components are characterized by power-law functions of the velocity, $\rho\propto v^{-m}$ for the inner ejecta and $\rho\propto v^{-n}$ for the outer ejecta. 
We assume that the slope $m$ for the inner ejecta is smaller than $3$ so that the mass of the inner ejecta does not diverge. 
By introducing a parameter $w_\mathrm{c}$, the location of the interface between the inner and outer ejecta in the velocity coordinate is specified as $v=w_\mathrm{c}v_\mathrm{ej}$. 
Therefore, the density profile is described as follows,
\begin{equation}
\rho(t,R)=\left\{\begin{array}{ccl}
\frac{f_\mathrm{3}M_\mathrm{ej}}{4\pi \omega_\mathrm{c}^3v_\mathrm{ej}^3t^3}\left(\frac{R}{w_\mathrm{c}v_\mathrm{ej}t}\right)^{-m}&\mathrm{for}&R\leq w_\mathrm{c}v_\mathrm{ej}t,\\
\frac{f_\mathrm{3}M_\mathrm{ej}}{4\pi \omega_\mathrm{c}^3v_\mathrm{ej}^3t^3}\left(\frac{R}{w_\mathrm{c}v_\mathrm{ej}t}\right)^{-n}&\mathrm{for}&w_\mathrm{c}v_\mathrm{ej}t<R\leq v_\mathrm{ej}t,\\
0,&\mathrm{for}&v_\mathrm{ej}t<R,
\end{array}\right.
\label{eq:ejecta_density}
\end{equation}
with a numerical factor $f_{l}$ given by,
\begin{equation}
f_l=\frac{(n-l)(l-m)}{n-m-(l-m)w_\mathrm{c}^{n-l}}.
\end{equation}
The integration of $4\pi r^2\rho(t,R)$ with respect to the radius $R$ from $0$ to $vt$ gives the mass $M(v)$ of the ejecta travelling at velocities slower than $v$,
\begin{eqnarray}
M(v)&=&\int_0^{vt}4\pi R^2\rho(t,R)dR
\nonumber\\
&=&
\left\{\begin{array}{llc}
\frac{f_\mathrm{3}M_\mathrm{ej}}{3-m}\left(\frac{v}{w_\mathrm{c}v_\mathrm{ej}}\right)^{3-m}
&\mathrm{for}&v\leq w_\mathrm{c}v_\mathrm{ej},\\
\frac{f_\mathrm{3}M_\mathrm{ej}}{3-n}
\left[\left(\frac{v}{w_\mathrm{c}v_\mathrm{ej}}\right)^{3-n}
-\frac{n-m}{3-m}
\right]
&\mathrm{for}&w_\mathrm{c}v_\mathrm{ej}<v.\\
\end{array}\right.
\end{eqnarray}
Thus, he masses, $M_\mathrm{inner}$ and $M_\mathrm{outer}$, of the inner and outer ejecta are given by
\begin{equation}
M_\mathrm{inner}=\frac{n-3}{n-m-(3-m)w_\mathrm{c}^{n-3}}M_\mathrm{ej},
\end{equation}
and
\begin{equation}
M_\mathrm{outer}=\frac{3-m-(3-m)w_\mathrm{c}^{n-3}}{n-m-(3-m)w_\mathrm{c}^{n-3}}M_\mathrm{ej}.
\end{equation}

In a similar way, the kinetic energy of the ejecta slower than $v$ is obtained as follows,
\begin{eqnarray}
E(v)&=&2\pi\int^{vt}_0\rho(t,R)v^2R^2dR
\nonumber\\
&=&\left\{\begin{array}{ccl}
\frac{f_\mathrm{3}M_\mathrm{ej}v_\mathrm{ej}^2\omega_\mathrm{c}^2}{2(5-m)}
\left(\frac{v}{\omega_\mathrm{c}v_\mathrm{ej}}\right)^{5-m}
&\mathrm{for}&v\leq \omega_\mathrm{c}v_\mathrm{ej},
\\
\frac{f_\mathrm{3}M_\mathrm{ej}v_\mathrm{ej}^2\omega_\mathrm{c}^2}{2(5-n)}
\left[
\left(\frac{v}{\omega_\mathrm{c}v_\mathrm{ej}}\right)^{5-n}
-\frac{n-m}{5-m}\right]
&\mathrm{for}&\omega_\mathrm{c}v_\mathrm{ej}<v.
\end{array}\right.
\end{eqnarray}
The total kinetic energy of the ejecta is given by $E_\mathrm{sn}=E(v_\mathrm{ej})$. 
For a small $w_\mathrm{c}$ and a large $n$, the numerical constants $f_3$ and $f_5$ weakly depend on the value of $w_\mathrm{c}$. 
The break velocity $w_\mathrm{c}v_\mathrm{ej}$ dividing the inner and outer parts of the ejecta is obtained for a given set of the mass $M_\mathrm{tot}$, the kinetic energy $E_\mathrm{sn}$, and the parameters specifying the density structure of the ejecta as follows,
\begin{equation}
w_\mathrm{c}v_\mathrm{ej}=\sqrt{
\frac{2f_\mathrm{5}E_\mathrm{sn}}{f_\mathrm{3}M_\mathrm{ej}}
}.
\end{equation}
We note that the break velocity does not depend on the parameters $m$ and $n$ so much, as long as $w_\mathrm{c}$ is small. 
Thus, the ratio between the mass $M_\mathrm{tot}$ and the kinetic energy $E_\mathrm{sn}$ are the dominant factor determining the velocity of the expansion. 

The gas in the ejecta is initially assumed to be cold, i.e., the pressure is negligibly small. 
We carry out the following semi-analytic and numerical calculations by assuming that the ejecta is still tightly coupled with radiation and thus adopt an ideal gas equation of state with an adiabatic index of $4/3$. 
In the following, we set $E_\mathrm{sn}=10^{51}$ erg and $M_\mathrm{ej}=10\ M_\odot$.

\subsection{Injection of Relativistic Gas}
After the formation of the freely expanding ejecta, a central compact remnant, a new-born neutron star or a black hole, starts depositing energy into the ejecta via some mechanism. 
For both scenarios, the energy is generally deposited in a region whose physical scale is much smaller than that of the ejecta. 
Thus, the energy density of the injected gas would soon be dominated by the kinetic energy as the gas expands even when it is initially dominated by the internal energy. 
We simply assume that the energy is deposited at a constant rate $L$ in a spherical manner. 
Specifically, we focus on the case in which the energy deposition is realized as an injection of relativistic gas. 
In other words, the internal energy of the injected gas is much larger than its rest-mass energy. 
The ratio $\Gamma_\mathrm{cr}$ of the energy injection rate to the mass injection rate $\dot{M}$,
\begin{equation}
\Gamma_\mathrm{cr}=\frac{L}{\dot{M}},
\end{equation}
is assumed to be larger than unity. 
In practice, we assume that the relativistic gas is uniformly injected within an injection radius $R_\mathrm{in}$, which is a small fraction of the physical scale of the ejecta. 
The initial velocity of the gas is set to zero. 
The injected gas radially expands at the expense of its internal energy.  
Thus, the kinetic energy of the gas soon dominates its total energy. 
As a result, the flow becomes highly relativistic at distances far from the injection radius. 
Since the energy flux of the flow is dominated by the kinetic one, the following relation between the density $\rho$, the velocity $v$, and the Lorentz factor $\Gamma=(1-v^2)^{-1/2}$ of the gas and the energy injection rate holds at $R$,
\begin{equation}
\frac{L}{4\pi R^2}=\rho\Gamma^2v.
\label{eq:rho_w}
\end{equation} 
In particular, when the Lorentz factor reaches the terminal value given by $\Gamma_\mathrm{cr}$, the density is inversely proportional to the square of the radius $\rho\propto R^{-2}$, corresponding to a simple steady wind solution. 

Furthermore, the timescale required for the total amount of the injected energy to reach the kinetic energy $E_\mathrm{sn}$ of the ejecta is defined as follows,
\begin{equation}
t_\mathrm{c}=\frac{E_\mathrm{sn}}{L}. 
\end{equation}
We use this timescale to normalise time $t$. 

\section{EVOLUTION OF SPHERICALLY SYMMETRIC SHELL}\label{spherical}
In this section, we consider the hydrodynamical interaction of the ejecta and the relativistic wind with spherical symmetry. 
The expanding relativistic gas immediately sweeps the innermost layer of the ejecta and creates a hot bubble surrounded by a geometrically thin shell. 
The bubble is composed of the shocked relativistic wind, while the shell is the ejecta swept and compressed by the forward shock propagation. 
The formation and expansion of the shell and the hot bubble have also been discussed in the context of the interaction between a pulsar wind nebula and supernova ejecta by earlier studies and there are several analytic studies focusing on its dynamical evolution \citep[e.g.][]{1971ApJ...164L..95O,1977ASSL...66...53C,1984ApJ...280..797C,1992ApJ...395..540C,1998ApJ...499..282J}. 
The total mass of the gas injected as a wind is usually much smaller than that of the supernova ejecta, leading to the reverse shock front with a radius much smaller than that of the contact surface separating the shocked wind and the shocked ejecta. 
Therefore, the bubble fills a considerable fraction of the volume surrounded by the contact surface. 
On the other hand, the forward shock front is close to the contact surface. 
Thus,  the supernova ejecta swept by the forward shock, which we refer to as the shell, becomes geometrically thin. 
In order for the energy injection to power supernova light curves, most of the additional energy well exceeding the explosion energy of the supernova should be deposited while the supernova ejecta are still tightly coupled to photons. 
Thus, the adiabatic index of $4/3$ would be appropriate instead of $5/3$, which is usually used for supernova remnants harbouring pulsar wind nebulae. 
In the following, we describe the dynamical evolution of the shell and the hot bubble partly based on these earlier works but we modify them to match our assumption of the relativistic gas injection. 

\subsection{Expanding Hot Bubble}
The relativistic wind is terminated by a reverse shock at $R=R_\mathrm{rs}$. 
Since we consider massive ejecta, $M_\mathrm{sn}>1M_\odot$, the average velocity of the ejecta cannot be relativistic even when an energy 10 times larger than the kinetic energy of the ejecta itself, $E_\mathrm{tot}=10^{52}$ erg, is deposited into the ejecta, $\sqrt{2E_\mathrm{tot}/M_\mathrm{ej}}<0.1c$. 
Thus, the reverse shock also travels at non-relativistic speeds. 

We denote the velocities of the reverse shock and the flow in the downstream of the reverse shock front by $v_\mathrm{rs}$ and $v_\mathrm{d}$. 
From the shock jump condition, these velocities and the upstream velocity $v_\mathrm{w}$ satisfy the following relation,
\begin{eqnarray}
&&\frac{\gamma}{\gamma-1}\Gamma_\mathrm{w}\Gamma_\mathrm{d}^2
(v_\mathrm{w}-v_\mathrm{d})(v_\mathrm{d}-v_\mathrm{rs})
\nonumber\\
&&\hspace{4em}=
\Gamma_\mathrm{w}(1-v_\mathrm{w}v_\mathrm{rs})-
\Gamma_\mathrm{d}(1-v_\mathrm{d}v_\mathrm{rs}),
\end{eqnarray}
where $\Gamma_\mathrm{w}=(1-v_\mathrm{w}^2)^{-1/2}$ and $\Gamma_\mathrm{d}=(1-v_\mathrm{d}^2)^{-1/2}$ are the Lorentz factors corresponding to the velocities, $v_\mathrm{w}$ and $v_\mathrm{d}$. 
Assuming that the upstream velocity is ultra-relativistic and much faster than the downstream velocity, $\Gamma_\mathrm{w}\gg \Gamma_\mathrm{d}$ and $v_\mathrm{w}\simeq 1$, and the shock velocity is much smaller than the downstream velocity, $v_\mathrm{rs}\ll v_\mathrm{d}$, the downstream velocity is found to be
\begin{equation}
v_\mathrm{d}\simeq \gamma-1=\frac{1}{3}.
\label{eq:beta_rd}
\end{equation}
The pressure $p_\mathrm{d}$ of the post-shock gas is also obtained from the shock jump condition,
\begin{equation}
p_\mathrm{d}=\rho_\mathrm{w}\Gamma_\mathrm{w}^2
\frac{(v_\mathrm{w}-v_\mathrm{d})(v_\mathrm{w}-v_\mathrm{rs})}
{1-v_\mathrm{d}v_\mathrm{rs}},
\end{equation}
where $\rho_\mathrm{w}$ is the density of the relativistic wind at the reverse shock. 

When the reverse shock radius is much larger than the injection radius, $R_\mathrm{rs}\gg R_\mathrm{in}$, the total energy of the relativistic gas is dominated by its kinetic energy at the reverse shock front and thus the relation (\ref{eq:rho_w}) can be used. 
Then, the post-shock pressure leads to
\begin{equation}
p_\mathrm{d}\simeq \frac{(2-\gamma)L}{4\pi R_\mathrm{rs}^2v_\mathrm{w}}.
\end{equation}
We should note that this expression depends very weakly on the Lorentz factor of the wind as long as a highly relativistic wind velocity, $v_\mathrm{w}\simeq 1$, is assumed. 

In a similar way, the density $\rho_\mathrm{d}$ of the post-shock gas is obtained as follows,
\begin{equation}
\rho_\mathrm{d}\simeq\rho_\mathrm{w}
\frac{\Gamma_\mathrm{w}v_\mathrm{w}}
{\Gamma_\mathrm{d}v_\mathrm{d}}\simeq
\frac{\sqrt{\gamma(2-\gamma)}L}{4\pi (\gamma-1)R_\mathrm{rs}^2\Gamma_\mathrm{w}v_\mathrm{w}}.
\end{equation} 
Therefore, the internal energy density dominates over the rest-mass energy density, $p_\mathrm{d}/\rho_\mathrm{d}\sim \Gamma_\mathrm{w}$. 

Finally, we derive the temporal evolution of the pressure averaged over the shocked region. 
The internal energy injected into the shocked gas through the reverse shock front per unit time is the product of the surface area $4\pi R_\mathrm{rs}^2$ and the internal energy flux $p_\mathrm{d}v_\mathrm{d}/(\gamma-1)$.
On the other hand, the shocked gas loses its internal energy by adiabatic cooling. 
Therefore, from the first law of thermodynamics, one obtains the following differential equation governing the temporal evolution of the internal energy $E_\mathrm{th}$ in the shocked region,
\begin{equation}
\frac{dE_\mathrm{th}}{dt}=
(2-\gamma)L
-\frac{(\gamma-1)E_\mathrm{th}}{V_\mathrm{c}}\frac{dV_\mathrm{c}}{dt},
\label{eq:dE_th}
\end{equation}
where $V_\mathrm{c}$ is the volume of the shocked gas,
\begin{equation}
V_\mathrm{c}=\frac{4\pi}{3}(R_\mathrm{c}^3-R_\mathrm{rs}^3),
\end{equation}
and $R_\mathrm{c}$ is the radius of the contact discontinuity. 
Here we assume that the volume of the unshocked wind is much smaller than that of the region surrounded by the contact discontinuity, $R_\mathrm{rs}^3\ll R_\mathrm{c}^3$, and approximate the volume $V_\mathrm{c}$ as that of a sphere with the radius $R_\mathrm{c}$, $V_\mathrm{c}\simeq 4\pi R_\mathrm{c}^3/3$. 
Thus, the first law of thermodynamics is rewritten as follows,
\begin{equation}
\frac{d}{dt}\left[E_\mathrm{th}R_\mathrm{c}^{3(\gamma-1)}\right]=(2-\gamma)LR_\mathrm{c}^{3(\gamma-1)}.
\label{eq:dEdt}
\end{equation}
Assuming that $R_\mathrm{c}$ is proportional to a power of time $t$ with an exponent $\alpha$, $R_\mathrm{c}\propto t^\alpha$, Equation (\ref{eq:dEdt}) can be integrated as follows,
\begin{equation}
E_\mathrm{th}=\frac{2-\gamma}{1+3\alpha(\gamma-1)}Lt.
\label{eq:E_th}
\end{equation}
Thus, the average pressure in the shocked region is found to be,
\begin{equation}
p_\mathrm{b}=\frac{3(\gamma-1)E_\mathrm{th}}{4\pi R_\mathrm{c}^3}=
\frac{3(\gamma-1)(2-\gamma)}{1+3\alpha(\gamma-1)}\frac{Lt}{4\pi R_\mathrm{c}^3}. 
\label{eq:pb}
\end{equation}

\subsection{Self-similar Solution}
The profiles of hydrodynamical variables, the velocity, the density, and the pressure of the gas in the shocked ejecta are well described by the self-similar solutions presented by \cite{1977ASSL...66...53C,1984ApJ...280..797C} and \cite{1998ApJ...499..282J}, who considered the propagation of a strong shock wave in freely expanding spherical ejecta with power-law density profiles. 
Appendix \ref{sec:self_similar} provides the derivation of the solutions in detail. 

The radii of the forward shock and the contact discontinuity evolve according to the same dependence of time $t$,
\begin{equation}
R_\mathrm{fs}=At^\alpha,
\end{equation}
and
\begin{equation}
R_\mathrm{c}=\xi_\mathrm{c}At^\alpha,
\label{eq:R_c}
\end{equation}
where the exponent $\alpha$ is determined from the time dependence of the pressure of the hot bubble and expressed in terms of the exponent $m$ of the power-law density profile,
\begin{equation}
\alpha=\frac{6-m}{5-m}.
\end{equation}
The normalisation constant $A$ is determined so that the pressure of the solution at the contact discontinuity is equal to that of the hot bubble $p_\mathrm{b}$,
\begin{equation}
A=\left\{\frac{3(\gamma-1)(2-\gamma)L}{\alpha^2\gamma[1+3\alpha(\gamma-1)] \xi_\mathrm{c}^3\eta_\mathrm{c}(w_\mathrm{c}v_\mathrm{ej})^{m-3}f_\mathrm{3}M_\mathrm{ej}}\right\}^{1/(5-m)}.
\end{equation}
The two dimensionless constants $\xi_\mathrm{c}$ and $\eta_\mathrm{c}$ appearing in these expressions are determined by numerically solving the dimensionless equations of the self-similar solution. 
The former $\xi_\mathrm{c}$ gives the ratio of the radius of the contact discontinuity to that of the forward shock, $\xi_\mathrm{c}=R_\mathrm{c}/R_\mathrm{fs}$ and its value is slightly smaller than unity, reflecting that the shell is geometrically thin. 
The latter $\eta_\mathrm{c}$ stands for the value of the dimensionless pressure at the contact discontinuity. 
The numerical values for $m=0$, $1$, and $2$ are presented in Table \ref{table:self_similar}. 

\begin{table}
\begin{center}
  \caption{Dimensionless Constants Derived from the Self-similar Solution}
\begin{tabular}{cccc}
\hline\hline
m&$\alpha$&$\xi_\mathrm{c}$&$\eta_\mathrm{c}$\\
\hline
$0$&$1.2$&$0.9875$&$0.09993$\\
$1$&$1.25$&$0.9849$&$0.1706$\\
$2$&$1.333$&$0.9810$&$0.3931$\\
\hline\hline
\end{tabular}
\label{table:self_similar}
\end{center}
\end{table}

\begin{table*}
\begin{center}
  \caption{Numerical Values Characterizing the Supernova Ejecta and the Breakout Time}
\begin{tabular}{rrrrrrrrrrrr}
\hline\hline
$w_\mathrm{c}$&$m$&$n$&$f_3$&$f_5$&$f_\mathrm{br}$&$w_\mathrm{c}$&$m$&$n$&$f_3$&$f_5$&$f_\mathrm{br}$\\
\hline
     0.1&       0&       9&       2&   2.222&    2.71&     0.3&       0&       9&       2&   2.232&   2.722\\
     0.1&       0&      10&     2.1&     2.5&   3.049&     0.3&       0&      10&     2.1&   2.503&   3.052\\
     0.1&       0&      11&   2.182&   2.727&   3.326&     0.3&       0&      11&   2.182&   2.728&   3.327\\
     0.1&       0&      12&    2.25&   2.917&   3.557&     0.3&       0&      12&    2.25&   2.917&   3.557\\
     0.1&       1&       9&     1.5&       2&   4.585&     0.3&       1&       9&     1.5&   2.008&   4.603\\
     0.1&       1&      10&   1.556&   2.222&   5.094&     0.3&       1&      10&   1.556&   2.225&   5.099\\
     0.1&       1&      11&     1.6&     2.4&   5.501&     0.3&       1&      11&     1.6&   2.401&   5.503\\
     0.1&       1&      12&   1.636&   2.545&   5.835&     0.3&       1&      12&   1.636&   2.546&   5.835\\
     0.1&       2&       9&  0.8571&   1.714&   10.55&     0.3&       2&       9&  0.8572&    1.72&   10.59\\
     0.1&       2&      10&   0.875&   1.875&   11.54&     0.3&       2&      10&   0.875&   1.877&   11.55\\
     0.1&       2&      11&  0.8889&       2&   12.31&     0.3&       2&      11&  0.8889&       2&   12.32\\
     0.1&       2&      12&     0.9&     2.1&   12.93&     0.3&       2&      12&     0.9&     2.1&   12.93\\

\hline\hline
\end{tabular}
 \label{table:f}
\end{center}
\end{table*}
\subsubsection{Breakout of Hot Bubble}
The self-similar solution can only be applied until the forward shock reaches the interface separating the inner and outer ejecta. 
After the emergence of the forward shock from the interface, the forward shock accelerates according to the steep density gradient of the outer ejecta and therefore the assumption of the uniform pressure throughout the whole reverse shocked region is not justified. 
The time $t_\mathrm{br}$ when the shock front reaches the interface at $r=w_\mathrm{c}v_\mathrm{ej}t$ is obtained by equating the positions of the shock front and the interface at $t=t_\mathrm{br}$, $At_\mathrm{br}^\alpha=w_\mathrm{c}v_\mathrm{ej}t_\mathrm{br}$, which yields
\begin{equation}
t_\mathrm{br}=\left(\frac{w_\mathrm{c}v_\mathrm{ej}}{A}\right)^{5-m}=f_\mathrm{br}t_\mathrm{c},
\end{equation}
with
\begin{equation}
f_\mathrm{br}=
\frac{2\alpha^2\gamma[1+3\alpha(\gamma-1)] \xi_\mathrm{c}^3\eta_\mathrm{c}f_5}{3(\gamma-1)(2-\gamma)}.
\label{eq:f_br}
\end{equation}
This dimensionless coefficient depends only on the free parameters specifying the structure of the ejecta, $\omega_\mathrm{c}$, $m$, and $n$. 
The numerical values for several sets of the free parameters are presented in Table \ref{table:f}. 
Furthermore, the energy having been deposited until $t=t_\mathrm{br}$ is given by $Lt_\mathrm{br}=f_\mathrm{br}E_\mathrm{sn}$ and proportional to the kinetic energy of the ejecta. 
Thus, as long as the total energy $E_\mathrm{tot}$ supposed to be injected from the central engine is larger than $Lt_\mathrm{br}$ and the coupling between gas and radiation in the ejecta remains strong to prevent the injected energy leaking as radiation, the blast wave can always reach the interface while the energy injection is still ongoing. 

Taking ejecta with $m=1,\ n=10$, and $w_\mathrm{c}=0.1$ for example, the forward shock reaches the interface at 
\begin{equation}
t_\mathrm{br}=5.1t_\mathrm{c},
\end{equation}
and accordingly the total injected energy until the breakout amounts to about five times larger than the kinetic energy of the ejecta,
\begin{equation}
Lt_\mathrm{br}=5.1E_\mathrm{sn}.
\end{equation}

\section{HYDRODYNAMIC SIMULATION}\label{multi-D}
The dynamical evolutions of the ejecta, the hot bubble, and the relativistic wind after the breakout are difficult to treat in analytic ways. 
Thus, we carry out two-dimensional hydrodynamic simulations to reveal how they evolve and how the injected energy is distributed throughout the ejecta. 
We use a two-dimensional special relativistic hydrodynamics code developed by one of the authors. 
The simulation employs cylindrical coordinates $(r,z)$. 
The numerical code solves equations governing the temporal evolution of the density $\rho$, the velocity components $v_r$ and $v_z$ along the $r$- and $z$-axes, and the pressure $p$ of the gas. 
The governing equations are given by
\begin{equation}
\frac{\partial (\rho\Gamma)}{\partial t}+\frac{1}{r}\frac{\partial }{\partial r}(r\rho\Gamma v_r)+\frac{\partial}{\partial z}(\rho\Gamma v_z)=0,
\end{equation}
\begin{equation}
\frac{\partial (\rho h\Gamma^2 v_r)}{\partial t}+\frac{1}{r}\frac{\partial }{\partial r}(r\rho h\Gamma^2 v_r^2)+\frac{\partial p}{\partial r}+\frac{\partial}{\partial z}(\rho h\Gamma^2 v_rv_z)=0,
\end{equation}
\begin{equation}
\frac{\partial (\rho h\Gamma^2 v_z)}{\partial t}+\frac{1}{r}\frac{\partial }{\partial r}(r\rho h\Gamma^2 v_rv_z)+\frac{\partial}{\partial z}(\rho h\Gamma^2 v_z^2+p)=0,
\end{equation}
and
\begin{equation}
\frac{\partial (\rho h\Gamma^2)}{\partial t}+\frac{1}{r}\frac{\partial }{\partial r}(r\rho h\Gamma^2 v_r)+\frac{\partial}{\partial z}(\rho h\Gamma^2 v_z)=0,
\end{equation}
where the Lorentz factor $\Gamma$ and the specific enthalpy $h$ are given by
\begin{equation}
\Gamma=\frac{1}{\sqrt{1-v_r^2-v_z^2}},
\end{equation}
and
\begin{equation}
h=1+\frac{\gamma}{\gamma-1}\frac{p}{\rho}.
\end{equation}
Appendix \ref{sec:code} briefly describes the numerical method to solve these equations. 

\subsection{Adaptive Mesh Refinement}
Our code is equipped with an adaptive mesh refinement technique (\citealt{1989JCoPh..82...64B}, see, Appendix \ref{sec:code}) to better resolve tiny structures expected to appear in later stages of the evolution of the ejecta. 
The coordinates $r$ and $z$ cover the ranges of $0\leq r\leq 4.8\times 10^{16}$ cm and $-4.8\times 10^{16}\ \mathrm{cm}\leq z\leq 4.8\times 10^{16}\ \mathrm{cm}$. 
The whole computational domain is covered by $128\times 256$ numerical cells at the lowest resolution i.e., the refinement level $0$. 
When an AMR block needs a finer resolution, four new blocks are generated to cover the coarse block, realizing a resolution finer by a factor of two than the coarse one. 
The maximum refinement level is initially set to $l_\mathrm{max}=10$, achieving the minimum resolved size of $\Delta r=\Delta z=3.66\times 10^{11}$ cm. 
The corresponding effective number of numerical cells covering the computational domain is $32768\times65536$. 

\subsection{Numerical Setups}
We consider ejecta with an energy of $E_\mathrm{sn}=10^{51}$ erg and a mass of $M_\mathrm{ej}=10M_\odot$. 
The energy injection rate $L$ is set to $L=10^{46}$ erg s$^{-1}$, which leads to a characteristic timescale of $t_\mathrm{c}=10^5 $ s. 
The initial time $t_0$ is set to $t_0=0.1t_\mathrm{c}$. 
The parameters specifying the density structure of the ejecta are chosen as follows, $w_\mathrm{c}=0.1$, $m=1$, and $n=10$, which leads to the break velocity of $w_\mathrm{c}v_\mathrm{ej}=3.8\times 10^8$ cm s$^{-1}$. 
The initial density and velocity distributions of the ejecta are given by Equations (\ref{eq:ejecta_velocity}) and (\ref{eq:ejecta_density}) with $t=t_0$ and $R=(r^2+z^2)^{1/2}$. 
The pressure of the ejecta is assumed to be sufficiently small so that it does not violate the assumption of free expansion. 
Thus, we assume that the pressure of the ejecta is $1\%$ of the local kinetic energy density,
\begin{equation}
p(t_0,R)=0.01\rho(t_0,R) v(t_0,R)^2.
\end{equation}
The energy and mass are deposited within the injection radius $R_\mathrm{in}=3\times 10^{12}$ cm. 
The energy injection lasts until a total energy of $E_\mathrm{tot}=10^{52}$ erg is deposited. 
In practice, the energy density and the mass density of the numerical cells inside the radius are increased at every time steps as follows,
\begin{equation}
\frac{\partial (\rho h\Gamma^2)}{\partial t}=\frac{L}{V_\mathrm{in}},
\end{equation}
and
\begin{equation}
\frac{\partial (\rho\Gamma)}{\partial t}=\frac{L}{\Gamma_\mathrm{cr} V_\mathrm{in}},
\end{equation}
where $V_\mathrm{in}$ is the total volume of the numerical cells inside the injection radius $R_\mathrm{in}$.  
The parameter $\Gamma_\mathrm{cr}$ specifies the baryon richness of the injected gas and is set to $\Gamma_\mathrm{cr}=20$ in our simulation. 
Thus, the rest-mass energy of the injected gas is much smaller than the internal energy. 

The simulation follows the evolution of the ejecta from $t=0.1t_\mathrm{c}$ to $t=20.0t_\mathrm{c}$. 
The energy injection is terminated at $t=10.1t_\mathrm{c}$. 
After the termination of the energy injection, we reduce the maximum refinement level by $1$ to $l_\mathrm{max}=9$, so that we can follow the evolution of the ejecta toward freely expanding stages at a reasonable numerical cost. 
The outermost layer of the ejecta is initially at $R=0.1v_\mathrm{ej}t_\mathrm{c}=3.8\times 10^{13}$ cm. 
The ejecta is surrounded by a static gas whose density is inversely proportional to the square of the radius. 
The collision of the ejecta into the ambient medium leads to the formation of a couple of shock waves, the forward and reverse shocks propagating into the ambient medium and the ejecta. 
However, the density of the ambient medium is assumed to be sufficiently small so that it does not significantly affect the dynamical evolution of the ejecta.  
Although the ambient gas is at rest in the simulation, it can be regarded as a steady wind from the progenitor star with a constant mass-loss rate $\dot{M}_\mathrm{a}$ and uniform velocity $v_\mathrm{a}$. 
The normalisation of the density of the ambient gas adopted in the simulation corresponds to a wind with $\dot{M}_\mathrm{a}=10^{-6}\ M_\odot$ yr$^{-1}$ for a wind velocity of $v_\mathrm{a}=10^3$ km s$^{-1}$.  

\section{RESULTS}
\subsection{Evolution of the Shell and Bubble}
\begin{figure*}[tbp]
\begin{center}
\includegraphics[scale=0.42,bb=0 0 1191 1134]{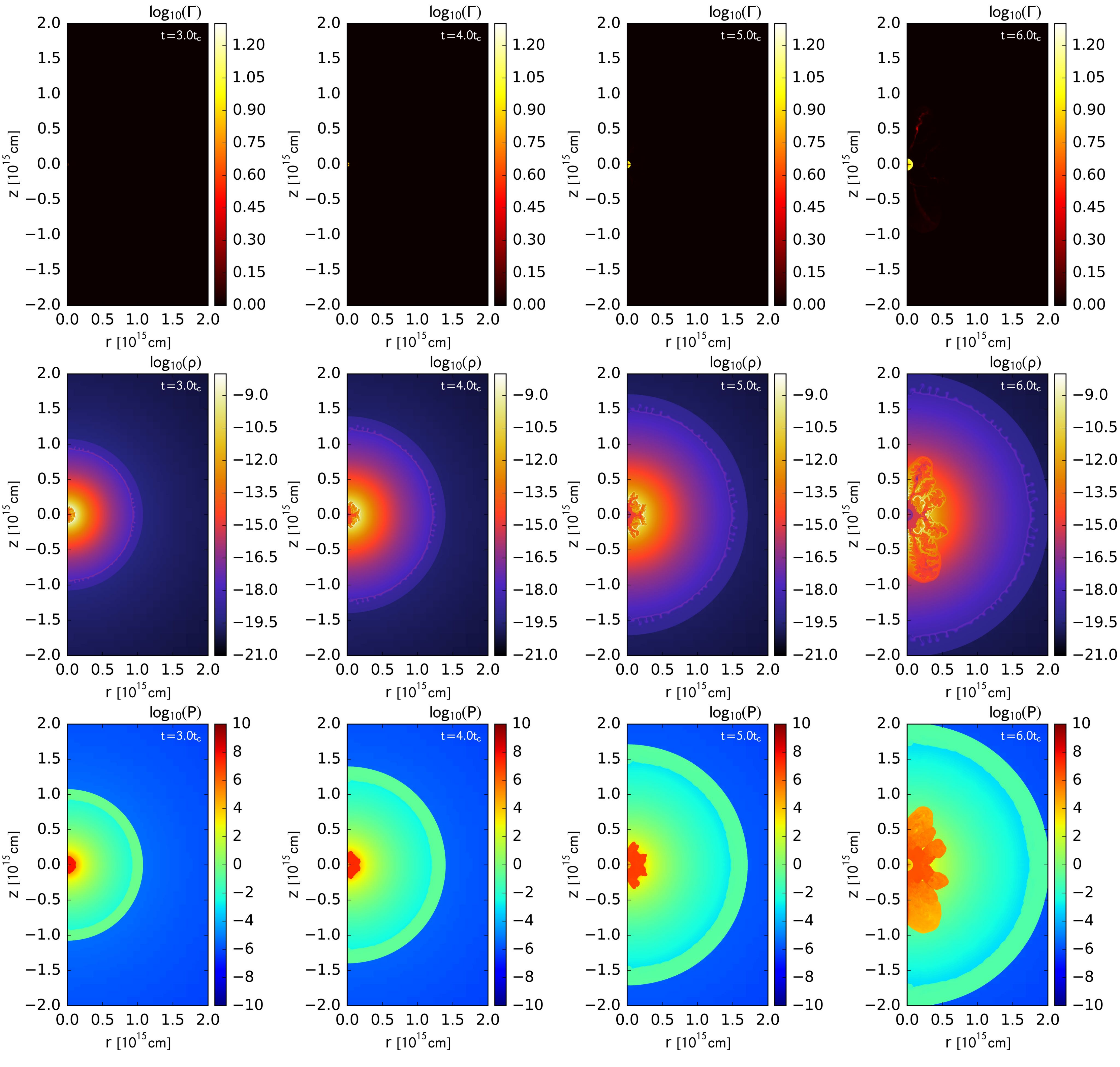}
\caption{Spatial distributions of the Lorenz factor (upper panels), the density (middle panels), and the pressure (bottom panels). 
The distributions at $t/t_\mathrm{c}=3.0$, $4.0$, $5.0$, and $6.0$ are shown from left to right. }
\label{fig:snap1}
\end{center}
\end{figure*}
\begin{figure*}[tbp]
\begin{center}
\includegraphics[scale=0.42,bb=0 0 1191 1134]{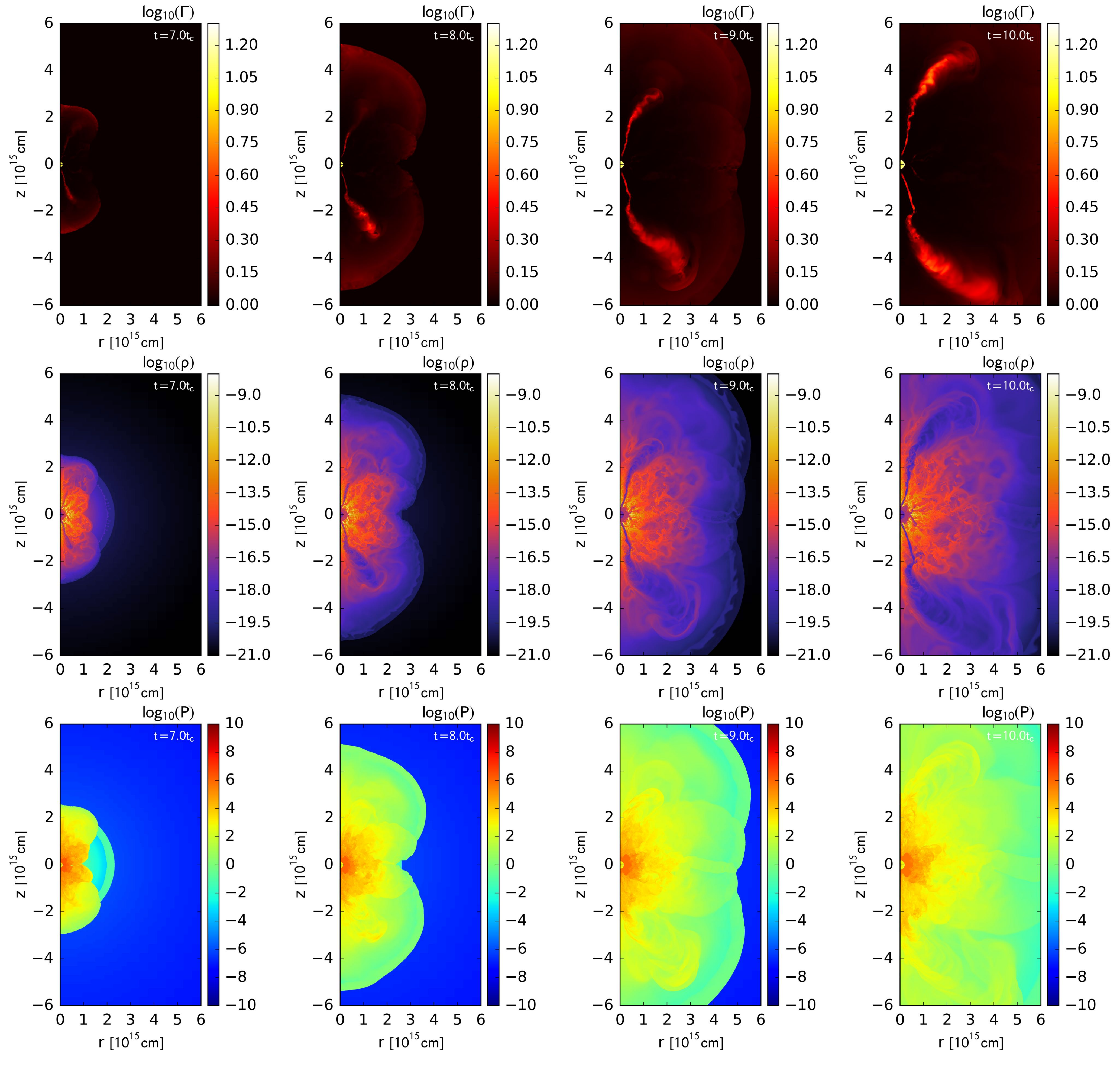}
\caption{Same as Figure \ref{fig:snap1}, but for $t/t_\mathrm{c}=7.0$, $8.0$, $9.0$, and $10.0$ from left to right. }
\label{fig:snap2}
\end{center}
\end{figure*}

\subsubsection{Self-similar Stage}
The spatial distributions of the Lorentz factor, the density, and the pressure of the ejecta at several epochs are shown in Figures \ref{fig:snap1} and \ref{fig:snap2}. 
First, we focus on the development and expansion of a hot bubble in the self-similar regime. 
At early stages, $t=3.0t_\mathrm{c}$ and $4.0t_\mathrm{c}$, the freely expanding ejecta and a hot bubble surrounded by a geometrically thin shell are clearly recognised. 
The shell is composed of the ejecta swept by the forward shock propagating in the ejecta. 
The contact discontinuity at $R=R_\mathrm{c}$ separates the shell and the shocked relativistic wind inside the shell. 
The hot bubble exhibits quite uniform pressure distributions as seen in the bottom panels of Figure \ref{fig:snap1}, which justifies our assumption in the self-similar analysis that the pressure of the hot bubble is given by averaging the thermal energy over the region inside the contact discontinuity, $R<R_\mathrm{c}$. 

One can already see the development of the Rayleigh-Taylor instability at the contact discontinuity, which stirs matter inside the shell. 
In fact, the development of the Rayleigh-Taylor instability is inevitable as long as the ejecta is pushed by the hot bubble. 
As we have seen in the previous section, the expanding hot bubble pushes and accelerates the shell, resulting in the contact surface growing faster than linear evolution, $R_\mathrm{c}\propto t^\alpha$ with $\alpha>1$. 
Therefore, in the rest frame of the accelerating shell, an inertial force acts on the shell toward the centre of the ejecta, i.e., it can be regarded as effective gravity. 
At the discontinuity separating the shocked wind and the shocked ejecta, denser media are stratified on top of dilute media and thus try to replace with the dilute ones according to the effective gravity. 
At this stage, however, the overall shape of the shell remains spherical. 

\subsubsection{Destruction of the Shell}
The dynamical evolution of the shell starts deviating from the one-dimensional picture at around $t=5.0t_\mathrm{c}$. 
The spatial distributions of the density and the pressure at $t=6.0t_\mathrm{c}$ show clear deviations from spherical symmetry. 
This is interpreted as leakage of the hot gas having been confined by the shell. 
The time of the destruction of the shell corresponds to the breakout time $t_\mathrm{br}=5.1t_\mathrm{c}$ at which the forward shock reaches the interface between the inner and outer ejecta. 

The reason why the hot bubble is well confined in the shell until this epoch is explained as follows. 
While the forward shock is still propagating in the inner ejecta, a large deviation from the spherical symmetry is not expected because of the shallow density gradient of the inner ejecta. 
As long as the exponent $m$ is smaller than $3$, the quantity $\rho R^3$, which has a dimension of mass, is an increasing function of the radius $R$. 
Therefore, a fluid element overshooting the shell would be subject to severe mass loading, resulting in deceleration of the fluid element. 
On the other hand, the density gradient of the outer ejecta is assumed to be very steep, reflecting that of the expanding envelope of the progenitor star.  
Thus, after the forward shock reaches the interface between the inner and outer ejecta, the confinement mechanism described above does not work. 
The steep density gradient of the outer ejecta allows the forward shock to efficiently accelerate, which results in an amplification of the displacement from the spherical shell. 
The acceleration of the forward shock and the amplified deviation from the spherical shell are seen in the spatial distributions of the density and the pressure at $t=6.0t_\mathrm{c}$ in Figure \ref{fig:snap1}. 

Our simulation demonstrates that the time of the transition from the quasi-spherical shell into the blowing off of the shell by the hot bubble can be predicted by the breakout time $t_\mathrm{br}=5.1t_\mathrm{c}$ obtained by the semi-analytic consideration. 

\begin{figure*}[tbp]
\begin{center}
\includegraphics[scale=0.6,bb=0 0 794 340]{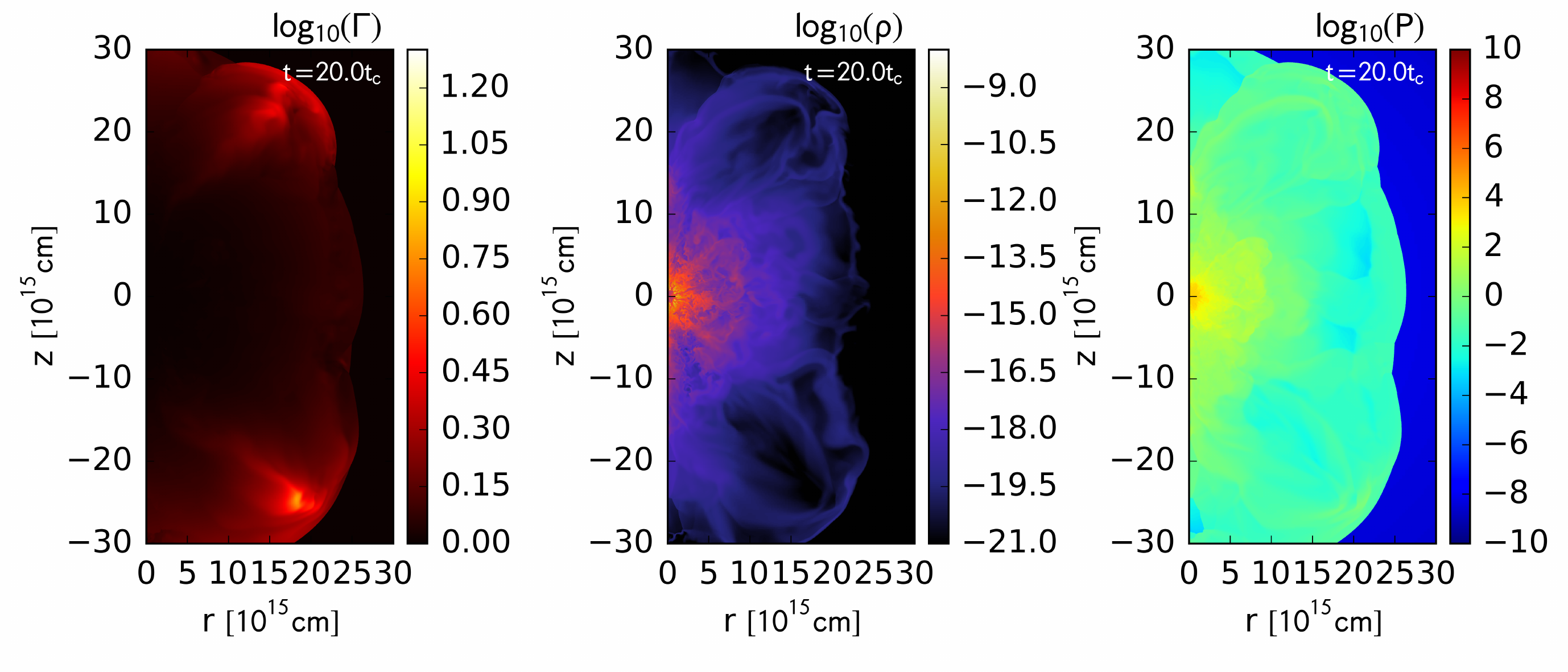}
\caption{Spatial distributions of the Lorentz factor (left), the density (middle), and the pressure (right) at $t=20t_\mathrm{c}$. }
\label{fig:snap_t20}
\end{center}
\end{figure*}

\subsubsection{Leaking Hot Bubble}\label{sec:hot_bubble}
After the quasi-spherical shell is destroyed, the forward shock driven by the pressure of the hot bubble immediately catches up with the outermost layer of the ejecta and finally expels the stratified layers. 
At that time, the pressure distribution within the hot bubble is no longer uniform as seen in the distribution at $t=6.0t_\mathrm{c}$ and $7.0t_\mathrm{c}$ in Figures \ref{fig:snap1} and \ref{fig:snap2}. 
This is because the forward shock now travels faster than the local sound speed of the gas in the central region. 
The gas immediately behind the forward shock front cannot communicate with the energy injection region via sound waves. 
Thus, the assumption of the uniform pressure distribution does not hold after the breakout. 
 
After the emergence of the hot bubble from the surface of the ejecta, the ejecta are powered by the pressure of the hot bubble, resulting in efficient mixing of the inner and outer ejecta and the relativistic wind. 
At later epochs, some low-density regions, which are filled with flows travelling at high Lorentz factors, $\Gamma\sim 5$, appear in the ejecta. 
These flows are composed of the relativistic gas having experienced the reverse shock and again accelerated at the expense of its internal energy. 
The maximum Lorentz factors of these relativistic outflows are roughly determined by the baryon richness of the injected relativistic gas, $\Gamma_\mathrm{cr}=20$.

\subsubsection{After the Termination of the Energy Injection}

After the total energy of $E_\mathrm{tot}=10^{52}$ erg has been deposited into the ejecta at $t=10.1t_\mathrm{c}$, the energy injection is terminated. 
The reverse shock in the relativistic wind gradually moves inward in the absent of the energy injection. 
The shock front finally reaches the centre of the ejecta and the whole injected gas is shocked. 

The spatial distributions of the Lorentz factor, the density, and the pressure at $t=20t_\mathrm{c}$ are shown in Figure \ref{fig:snap_t20}. 
At this time, the outermost layer of the ejecta reaches a distance of $\simeq 3\times 10^{16}$ cm from the centre. 
As a result of the reverse shock reaching the centre, the hole in the pressure distribution, which consists of unshocked relativistic wind and seen in the lower panels of Figure \ref{fig:snap2}, disappears. 
The disappearance of the unshocked relativistic wind can also be recognised in the density distribution in Figure \ref{fig:snap_t20}. 
However, the central region is still filled with relatively dilute gas and surrounded by a shell-like structure with higher densities. 
Highly relativistic flows emanating from the central region are now travelling in a distant region from the centre and they are gradually decelerating by colliding into surrounding gas.

\subsection{Comparison with Self-similar Solution}
\begin{figure}[tbp]
\begin{center}
\includegraphics[scale=0.55,bb=0 0 453 566]{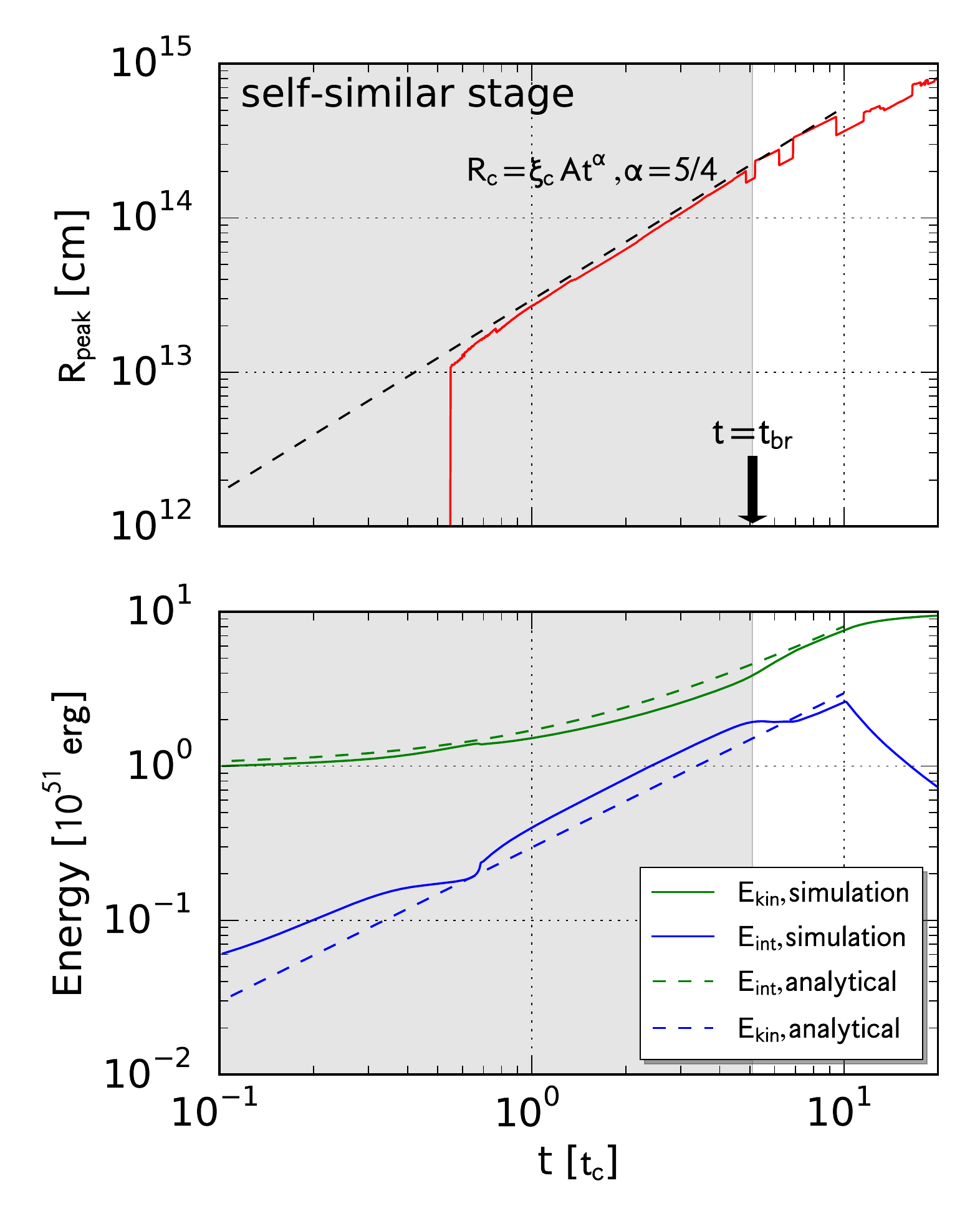}
\caption{Comparison between the self-similar solution and the numerical simulation. 
In the upper panel, the solid and dashed lines show the radius of the density peak in the ejecta in the simulation and the shell radius in the semi-analytic model. 
In the lower panel, the temporal evolutions of the kinetic and internal energies are shown. 
The solid lines correspond to those obtained in the simulation, while the dashed lines are predictions from the semi-analytic model. 
The shaded area represents the self-similar stage. }
\label{fig:evolution}
\end{center}
\end{figure}

\subsubsection{Self-similar Stage}
We quantitatively compare results of our simulation with the semi-analytic considerations assuming spherical symmetry. 
We identify the numerical cell with the highest density (referred to as the ``density peak'') and plot the radius of the density peak as a function of time in the upper panel of Figure \ref{fig:evolution}. 
This is compared with the temporal evolution of the radius of the contact discontinuity $R_\mathrm{c}$, Equation (\ref{eq:R_c}). 
The radius of the density peak well agrees with that of the semi-analytic estimation not only for the exponent but also for the normalisation $\xi_\mathrm{c}A$. 
This remarkable agreement proves that the semi-analytic approach based on the self-similar solution accurately captures the evolution of the quasi-spherical shell well confined in the inner ejecta. 

The lower panel of Figure \ref{fig:evolution} shows the temporal evolution of the kinetic energy $E_\mathrm{kin}$ and the internal energy $E_\mathrm{int}$.  
The kinetic energy of the ejecta is initially $E_\mathrm{kin}=E_\mathrm{sn}=10^{51}$ erg, while the internal energy is smaller than $10\%$ of the kinetic energy. 
After the simulation starts, the energy injection at the centre increases both the kinetic and internal energies. 
As we have shown, the self-similar solution predicts a linear increase in the internal energy, Equation (\ref{eq:E_th}). 
Since the total injected energy also increases in a linear manner, $E_\mathrm{tot}=Lt$, the conservation of the total energy means that the kinetic energy should evolve as follows,
\begin{equation}
E_\mathrm{kin}=E_\mathrm{sn}+\frac{(1+3\alpha)(\gamma-1)}{1+3\alpha(\gamma-1)}Lt,
\end{equation}
which is also a linear function of time. 
These temporal evolutions of the kinetic and internal energies are plotted as dashed lines in the lower panel of Figure \ref{fig:evolution}. 
The analytically obtained internal energy is slightly lower than that in the numerical simulation at the self-similar stage. 
This is because of the development of the Rayleigh-Taylor instability inside the hot bubble. 
As we have seen in the previous subsection, in the self-similar stage, the contact discontinuity is subject to mixing due to the Rayleigh-Taylor instability, while the overall shape of the shell keeps spherical symmetry. 
The mixing invokes downward motions of fluid elements and they dissipate their kinetic energies by colliding with each other. 
As a result of the additional heating, the semi-analytic model underestimates the internal energy in the numerical simulation. 

\subsubsection{After the Breakout}
In Figure \ref{fig:evolution}, the dashed lines are extrapolated to epochs after the breakout. 
After the breakout of the hot bubble from the shell, the numerically obtained kinetic and internal energies do not agree with the values obtained by the semi-analytic model. 
The internal energy temporally stops increasing immediately after the breakout. 
The evolution of the internal energy is governed by the balance between the supply from the relativistic wind through the reverse shock and adiabatic loss, Equation (\ref{eq:dE_th}). 
Since the energy injection from the central engine is still ongoing, the saturation of the internal energy indicates that the adiabatic loss becomes significant at this stage. 
This is also supported by the spatial distributions shown in Figure \ref{fig:snap2}. 
After the breakout, the ejecta blown by the hot bubble rapidly expand into the interstellar space. 
The expansion efficiently converts the internal energy of the gas into the kinetic energy, resulting in the temporal saturation of the internal energy. 

\subsubsection{After the Termination of the Energy Injection}
The internal energy of the ejecta rapidly decreases after $t=10.1t_\mathrm{c}$, because of the terminated energy injection. 
The internal energy has been supplied by the hot bubble through the reverse shock as long as the central engine is active. 
After the energy injection is terminated, the ejecta is no longer pushed by the hot bubble. 
Although collisions between fluid elements in the ejecta may contribute to the heating of the ejecta even after the termination of the energy injection, the contrubution is not sufficient to keep the internal energy growing. 
Thus, the internal energy continues to decrease due to adiabatic cooling. 
On the other hand, the kinetic energy continues to increase as the internal energy is lost.  
Although the decrease in the internal energy in our hydrodynamic simulation is due to adiabatic cooling, it can be lost via radiative diffusion. 
How seriously the radiative diffusion contributes to the energy loss and how much energy is expected to escape into the interstellar space as radiation will be discussed in the next section. 

\subsection{Radial Profiles}\label{sec:radial}
\begin{figure*}[tbp]
\begin{center}
\includegraphics[scale=0.4,bb=0 0 906 793]{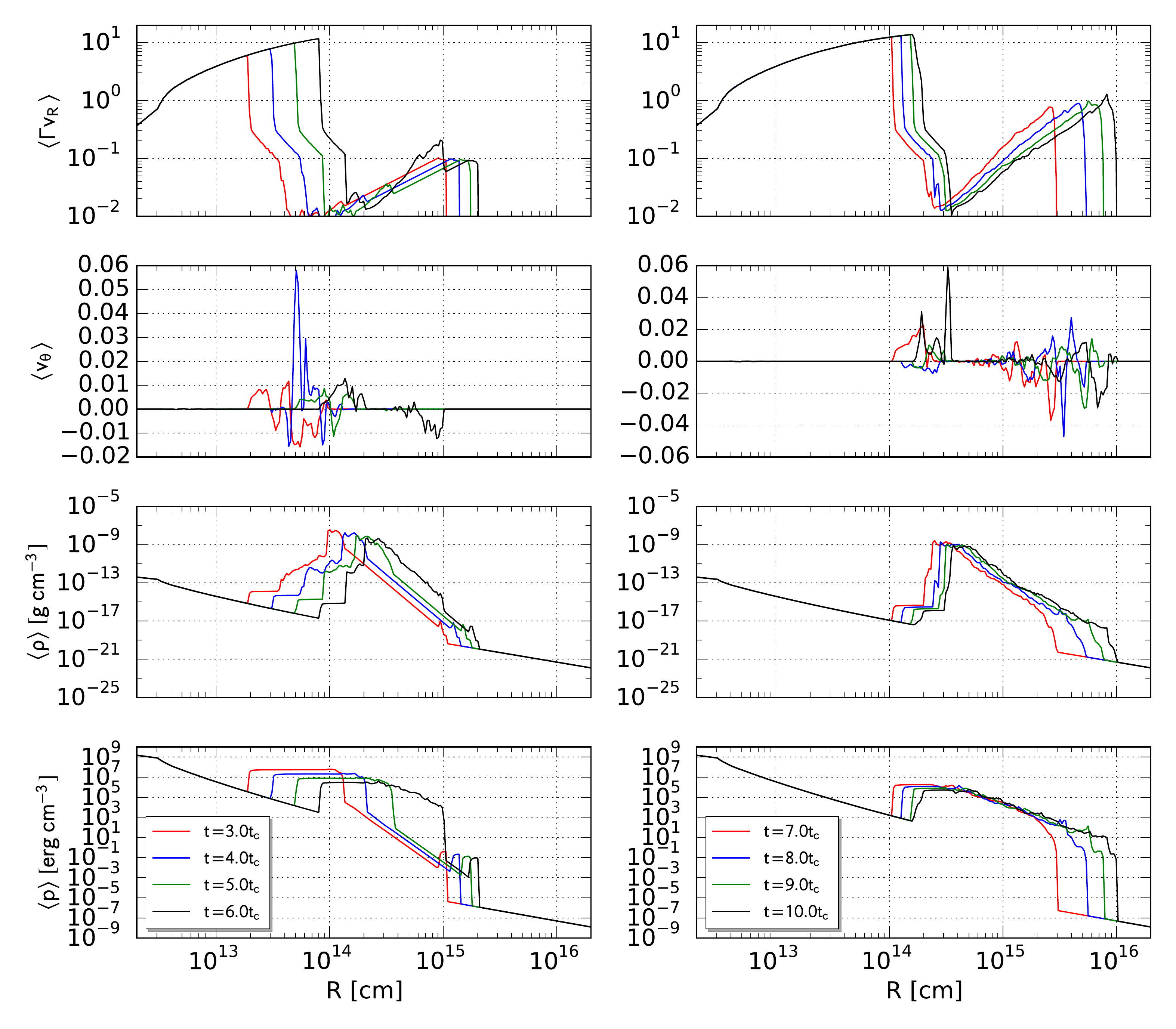}
\caption{Radial profiles of the 4-velocity along the radial direction $\langle\Gamma v_R\rangle$, the angular velocity $\langle v_\theta\rangle$, the density $\langle\rho\rangle$, and the pressure $\langle p\rangle$ from top to bottom. 
The left panels show the profiles at $t/t_\mathrm{c}=3.0$, $4.0$, $5.0$, and $6.0$, while those shown in the right panels are at $t/t_\mathrm{c}=7.0$, $8.0$, and $10.0$. }
\label{fig:radial}
\end{center}
\end{figure*}

We calculate angle-averaged radial profiles of physical quantities from snapshots of the simulation. 
We introduce spherical coordinates $(R,\theta)$, where $R$ is the distance from the centre $(r,z)=(0,0)$, $R=(r^2+z^2)^{1/2}$, and $\theta$ is the angle measured from the symmetry axis, $\theta=\arccos(z/R)$. 
We divide the numerical domain into a number of concentric shells with a width $\Delta R$ and calculate the following integrations for a quantity $Q$ over the shells at various distances,
\begin{equation}
\langle Q(R)\rangle=\frac{2\pi}{\Delta M}\int ^{R+\Delta R/2}_{R-\Delta R/2}\int^\pi_{0} Q\rho\Gamma r^2\sin\theta drd\theta,
\end{equation}
and
\begin{equation}
\Delta M=2\pi \int^{R+\Delta R/2}_{R-\Delta R/2}\int^\pi_{0}\rho\Gamma r^2\sin\theta drd\theta.
\end{equation}
In other words, the averaged values are mass-weighted. 

Figure \ref{fig:radial} shows the radial profiles of the 4-velocity along the radial direction $\langle\Gamma v_R\rangle$, the angular velocity $\langle v_\theta\rangle$, the density $\langle\rho\rangle$, and the pressure $\langle p\rangle$. 
We again confirm that the radial profile of the density is well represented by a geometrically thin shell and the pressure inside the hot bubble is almost uniform before the breakout. 
The radial profiles of the 4-velocity $\langle\Gamma v_R\rangle$ show that relativistic gas injected at the centre travels at highly relativistic speeds and decelerates at the reverse shock to a sub-relativistic speed, which well agrees with the analytic value, $v_{R}\simeq 1/3$, Equation (\ref{eq:beta_rd}). 

After the breakout, the hot bubble leaking from the shell starts accelerating the outer ejecta, resulting in an increase in the radial velocity. 
The 4-velocity of the outermost layer of the ejecta reaches $\Gamma v_\mathrm{R}\sim 1$. 
The profiles of the angle-averaged angular velocity suggest that angular velocities of most ejecta are much smaller than the radial velocities of the corresponding layers. 

\begin{figure}[tbp]
\begin{center}
\includegraphics[scale=0.5,bb=0 0 453 680]{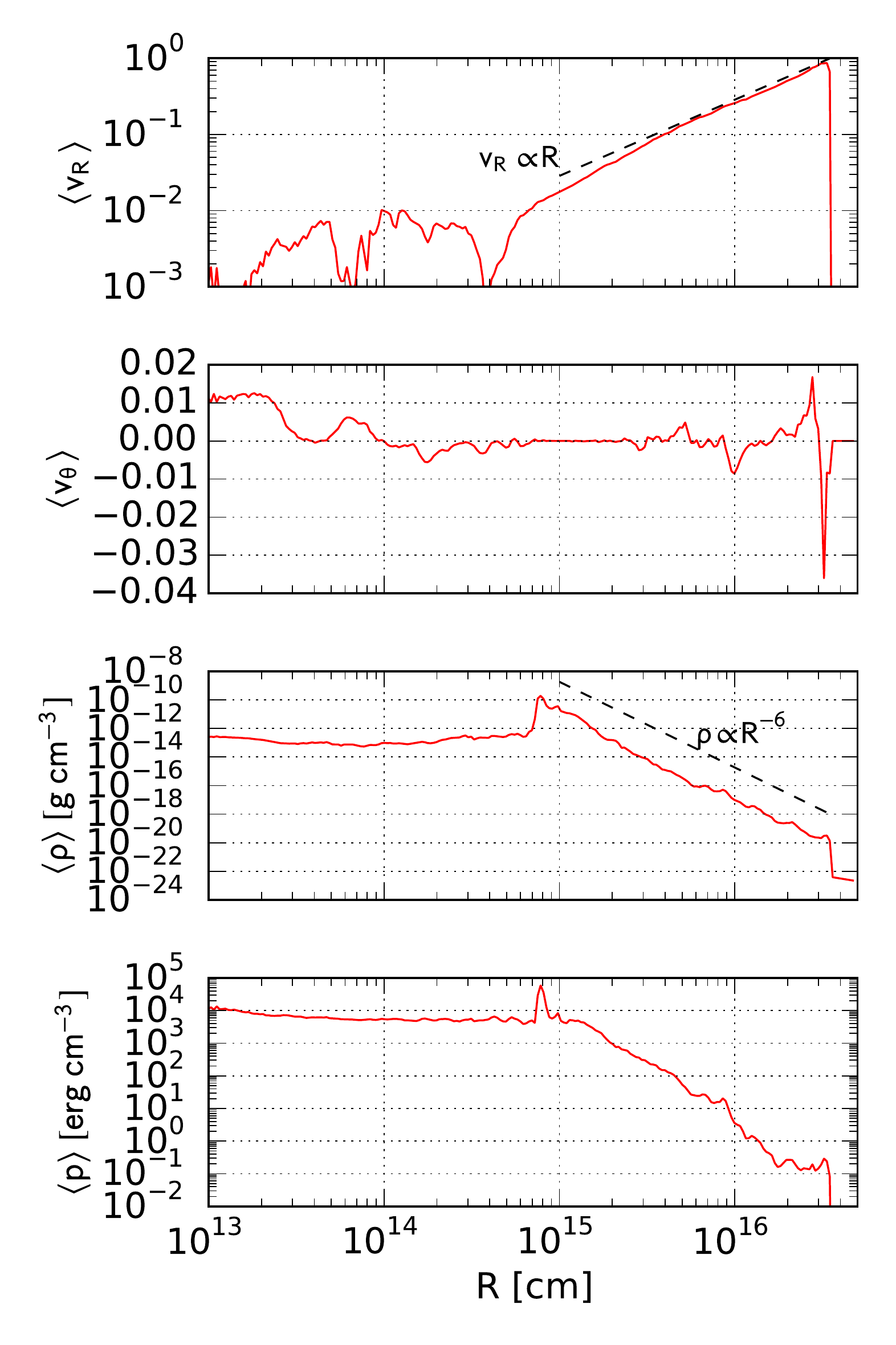}
\caption{Radial profiles of the radial velocity $\langle v_R\rangle$, the angular velocity $\langle v_\theta\rangle$, the density $\langle\rho\rangle$, and the pressure $\langle p\rangle$ at $t/t_\mathrm{c}=20.0$ from top to bottom. }
\label{fig:radial_final}
\end{center}
\end{figure}

Figure \ref{fig:radial_final} shows the radial profiles of the radial velocity $\langle v_R\rangle$, the angular velocity, the density and the pressure at $t=20.0t_\mathrm{c}$. 
Since the energy injection has been terminated, no relativistic wind is seen in the profiles. 
The dashed line in the top panel of Figure \ref{fig:radial_final} indicates a radial velocity profile proportional to the radius, $v_R\propto R$. 
The linear relation well agrees with the numerically obtained radial velocity profile, suggesting that the ejecta have already been freely expanding. 
In addition, the dashed line plotted along with the density profile is a power-law distribution with an exponent $-6$, $\rho\propto R^{-6}$. 
The density profile of the ejecta is well represented by the power-law function. 
Furthermore, the angular velocity is much smaller than the radial velocity at least in the power-law part of the ejecta. 
Thus, the density distribution of the ejecta in the velocity coordinate is also given by a power-law function with the exponent, $\rho\propto v^{-6}$, from $v\simeq 0.01$ to $v\simeq 1$.

\subsection{Mass and Energy Distributions}
\begin{figure*}[tbp]
\begin{center}
\includegraphics[scale=0.4,bb=0 0 906 680]{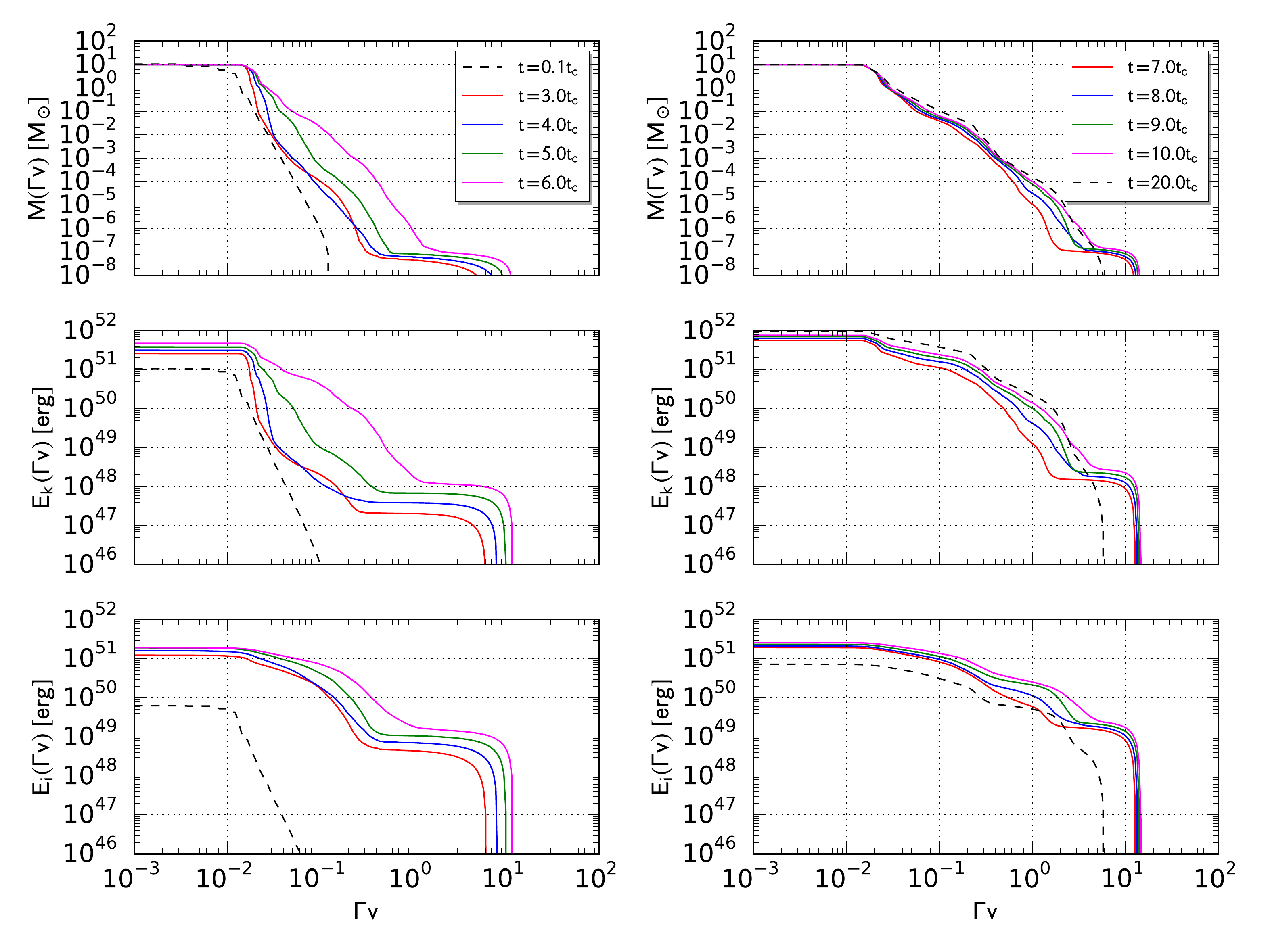}
\caption{Cumulative mass, kinetic energy, and internal energy distributions in the ejecta. from top to bottom. 
The left panels show the distributions at $t/t_\mathrm{c}=3.0$, $4.0$, $5.0$, and $6.0$, while those shown in the right panels are at $t/t_\mathrm{c}=7.0$, $8.0$, $9.0$, $10$, and $20.0$.
The dash-dotted lines in the left panels show the initial distributions.}
\label{fig:spec}
\end{center}
\end{figure*}

We define the following cumulative mass, kinetic energy, and internal energy distributions of the ejecta in order to quantify how much mass and energy are distributed in different parts of the ejecta,
\begin{equation}
M(\Gamma v)=2\pi \int_{>\Gamma v}\rho\Gamma rdrdz,
\end{equation}
\begin{equation}
E_\mathrm{k}(\Gamma v)=2\pi \int_{>\Gamma v}\rho\Gamma(\Gamma-1) rdrdz,
\end{equation}
and
\begin{equation}
E_\mathrm{i}(\Gamma v)=2\pi \int_{>\Gamma v}\left(\frac{\gamma}{\gamma-1} \Gamma^2-1\right)p rdrdz,
\end{equation}
where the integrations are carried out over fluid elements with 4-velocities larger than $\Gamma v$. 

The distributions at several epochs are presented in Figure \ref{fig:spec}. 
In each panel of Figure \ref{fig:spec}, one can recognise a component represented by an almost flat distribution, $\propto (\Gamma v)^0$ from $\Gamma v\sim1$ to $\Gamma v\sim 10$. 
This corresponds to the unshocked relativistic wind. 
On the other hand, in low-velocity regimes, $\Gamma v<0.01$, the distribution again shows a flat shape, which continuously connects with a steeply declining part at around $\Gamma v\sim 0.02$-$0.03$. 
These flat and steep parts represent the shell, which has a considerable fraction of the mass and the kinetic energy, and the outer ejecta surrounding the shell. 
The break velocity connecting the flat and steep parts gives a characteristic velocity of the shell, which reasonably agrees with $(2E_\mathrm{tot}/M_\mathrm{ej})^{1/2}\simeq 0.03$. 
After the breakout, the outer ejecta expelled by the hot bubble exhibit quite shallow kinetic energy distributions from $\Gamma v\sim 0.03$ to $\Gamma v\sim 1$, which means that a non-negligible fraction of the total energy has been deposited into the outer ejecta travelling at sub-relativistic speeds through the expansion of the hot bubble. 

In the right panels of Figure \ref{fig:spec}, the dashed lines show the mass and energy distributions at $t=20.0t_\mathrm{c}$. 
Since the energy injection has been terminated, the internal energy of the ejecta has been reduced due to adiabatic expansion. 
However, the overall shapes of the mass and kinetic energy distributions from $\Gamma v=0.01$ to $\Gamma v=1.0$ remain almost unchanged from that at $t=10.0t_\mathrm{c}$.  
 
\subsection{Photospheric Emission}
\begin{figure*}[tbp]
\begin{center}
\includegraphics[scale=0.6,bb=0 0 680 566]{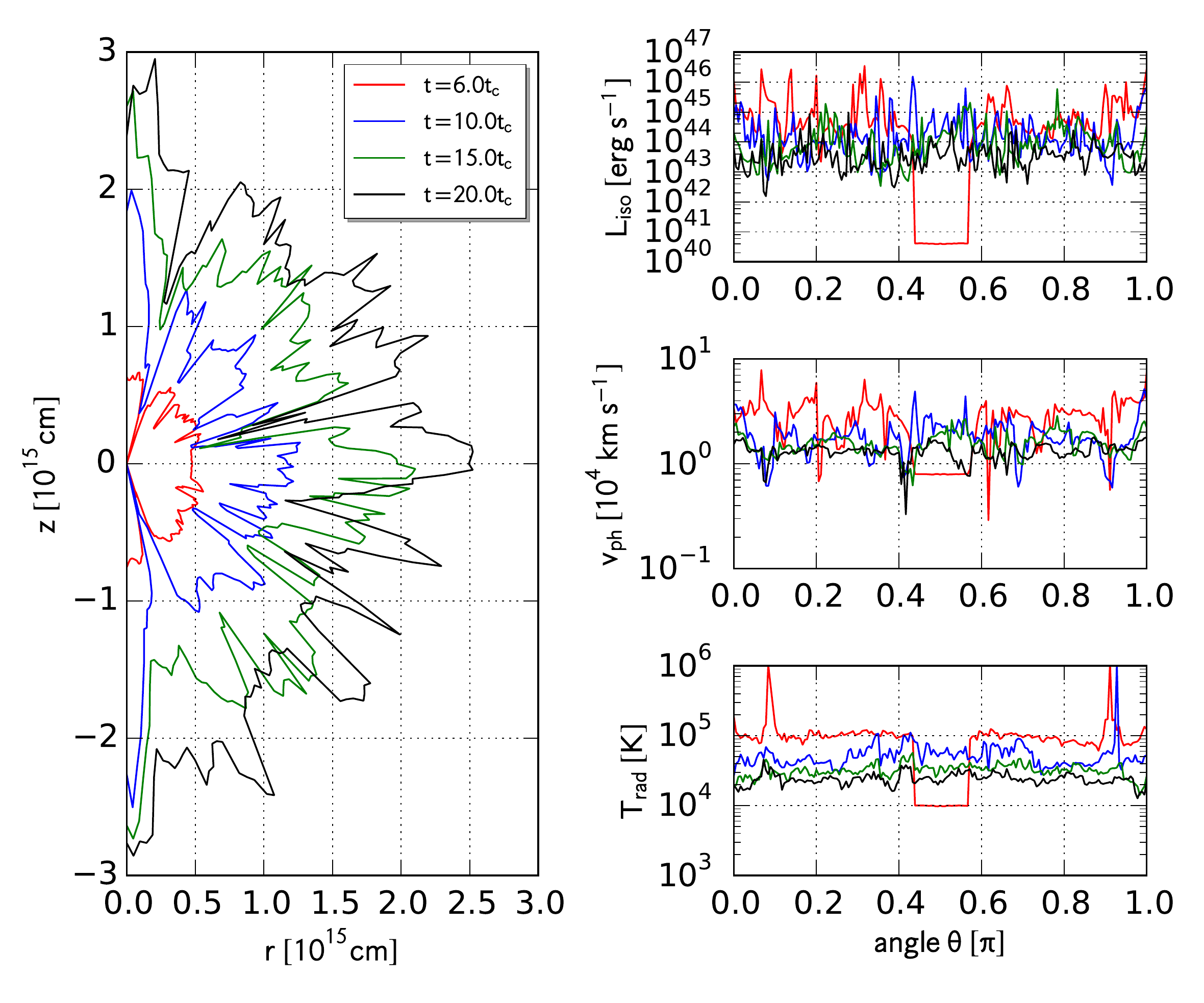}
\caption{Photosphere (left panel) and physical variables at the photosphere (right panels) calculated from several snapshots of the simulation. 
In each panel, solid lines show results obtained for $t/t_\mathrm{c}=6.0$, $10.0$, $15.0$, and $20.0$. 
In the right panels, the isotropic luminosity, the density, and the radiation temperature at the photosphere are shown as functions of the angle $\theta$. 
}
\label{fig:photo}
\end{center}
\end{figure*}

We use the spatial distributions of the physical variables obtained by our simulation to identify the photosphere located in the ejecta. 
The location of the photosphere generally depends on the viewing angle from which the ejecta are observed. 
We define the optical depth $\tau(t,R,\theta)$ from the outer boundary of the numerical domain to the centre along a radial direction with a fixed angle $\theta$,
\begin{equation}
\tau(t,R,\theta)=\kappa\int_R^{R_\mathrm{out}}\rho(t,R',\theta) dR',
\label{eq:tau}
\end{equation}
where $R_\mathrm{out}$ is the outer boundary radius. 
Then, we define the photospheric radius $R_\mathrm{ph}$ as the radius at which the optical depth is equal to unity, $\tau(t,R_\mathrm{ph},\theta)=1$. 
As a result, the radius is expressed as a function of time and the angle, $R_\mathrm{ph}(t,\theta)$. 
The opacity $\kappa$ is assumed to be a constant, $\kappa=0.1$ cm$^{2}$ g$^{-1}$, for simplicity. 

The rate of radiative energy loss through the photosphere is estimated in the following way. 
Differentiating the definition of the photospheric radius $\tau(t,R_\mathrm{ph},\theta)$, one obtains the time derivative of the photospheric radius,
\begin{equation}
\frac{dR_\mathrm{ph}}{dt}=-\left(\frac{\partial\tau}{\partial t}\right)\left(\frac{\partial\tau}{\partial R}\right)^{-1}=
-\frac{1}{\rho}\int^\infty_{R_\mathrm{ph}}\frac{\partial \rho}{\partial t}dR.
\end{equation}
From the continuity equation for spherically symmetric flows, Equation (\ref{eq:cont_spherical}), the above equation yields the following expression,
\begin{equation}
\frac{dR_\mathrm{ph}}{dt}=v_\mathrm{ph}-\frac{2}{\rho_\mathrm{ph}}\int^\infty_{R_\mathrm{ph}}\frac{\rho v_\mathrm{R}}{R}dR,
\label{eq:dRph/dt}
\end{equation}
where $v_\mathrm{ph}$ and $\rho_\mathrm{ph}$ are the velocity and the density at the photosphere $R=R_\mathrm{ph}$.

We assume that the internal energy of the ejecta is dominated by radiation. 
Thus, the radiation energy in the ejecta is given by $u_\mathrm{rad}=3p$. 
As given in Equation (\ref{eq:dRph/dt}), the expansion of the photosphere generally delays from the flow. 
Thus, the volume $\Delta V$ of the layer that becomes transparent to photons within a short duration from $t$ to $t+\Delta t$ yields,
\begin{equation}
\Delta V=4\pi R_\mathrm{ph}^2\left(v_\mathrm{ph}-\frac{dR_\mathrm{ph}}{dt}\right)\Delta t.
\end{equation}
Therefore, the radiation energy decoupling from the ejecta per unit time within the duration can be estimated to be
\begin{equation}
\frac{dE}{dt}=u_\mathrm{rad}\frac{\Delta V}{\Delta t}=4\pi R_\mathrm{ph}^2u_\mathrm{rad}\left(v_\mathrm{ph}-\frac{dR_\mathrm{ph}}{dt}\right).
\label{eq:Liso}
\end{equation}
This rate can be regarded as the isotropic luminosity $L_\mathrm{iso}$ of the photospheric emission when the light travelling time across the photospheric radius is much smaller than the timescale of the photospheric emission. 
In this case, the assumption of short light travelling times is marginally justified because the photospheric radii are $(5-20)\times10^{15}$ cm, which corresponds to light travelling times of the order of $10^{5}$ s, at $t\simeq 10t_\mathrm{c}= 10^{6}$ s. 
Thus, we use Equation (\ref{eq:Liso}) to estimate the isotropic luminosity of the photospheric emission. 
Furthermore, the radiation temperature at the photosphere can be estimated by $T_\mathrm{rad}=(3p_\mathrm{ph}/a_\mathrm{r})^{1/4}$, where $p_\mathrm{ph}$ is the pressure at the photosphere and $a_\mathrm{r}$ is the radiation constant. 

We calculate the photospheric radius and physical variables at the photosphere from several snapshots of our simulations. 
The photospheric radius is given as a function of the angle $\theta$. 
The left panel of Figure \ref{fig:photo} shows the photosphere at $t/t_\mathrm{c}=6.0$, $10.0$, $15.0$, and $20.0$. 
The right panels of Figure \ref{fig:photo} show the isotropic luminosity estimated by Equation (\ref{eq:Liso}), the radial velocity, and the radiation temperature at the photosphere as functions of the angle $\theta$. 
The hot bubble is still inside the photosphere until the breakout $t=5.1t_\mathrm{c}$. 
The photospheric radius after the breakout is a very complicated function of the angle $\theta$, reflecting the mixing of the ejecta by the hot bubble. 
Accordingly, the physical variables at the photosphere also exhibit significant fluctuations depending on the angle. 
We can expect photospheric emission with an average luminosity of the order of $\sim 10^{43-45}\ \mathrm{erg}\ \mathrm{s}^{-1}$ and radiation temperature of $(2-10)\times 10^4$ K when we assume blackbody spectra, which are consistent with peak luminosities and relatively blue spectra of SLSNe-I.

\section{DISCUSSIONS}\label{discussion}
In this section, we discuss properties of the supernova ejecta powered by the energy injection from the central compact object. 
\subsection{Density Structure of Supernova Ejecta}\label{sec:density_structure}
In Section \ref{sec:radial} and Figure \ref{fig:radial_final}, we have shown that the radial density profile of the ejecta is well described by a power-law function of the velocity $v$ with an exponent $-6$. 
This distribution is shallower than those of expanding envelopes in normal SNe (from $n=9$ to $n=12$ as assumed in our model), which is clearly a consequence of the additional energy injection. 
The density distribution of freely expanding ejecta is a key to distinguishing existing models of SLSNe and other extraordinary SNe. 
In the following, we discuss how ejecta with such density structure form in the presence of the central energy source. 
For simplicity, we restrict ourselves to the Newtonian limit. 
 
We consider the following idealistic case. 
In normal SNe, the strong blast wave driven by a point explosion is the only way to transport the explosion energy to the outermost layer. 
As a result, only a small fraction of the energy is deposited in the outer envelope, while the inner ejecta posses the bulk of the energy, leading to a steep kinetic energy distribution \citep{1999ApJ...510..379M}. 
In contrast to normal SNe, the central engine deposits an energy much larger than the supernvova ejecta for a timescale much longer than the expansion timescale of the ejecta. 
In our setting, the additional energy is injected at a constant rate $L$. 
Therefore, if the energy is continuously transferred throughout all the layers of the ejecta without loss or stagnation, the kinetic energy flux $\rho v^3$ of the ejecta powered by the energy injection at a radius $R_0$ roughly proportional to the energy flux $L/(4\pi R_0^2)$. 
The presence of relativistic flows leaking from the hot bubble makes the efficient energy transport possible. 
The flows can directly bring and deposit the additional energy throughout different layers. 
Thus, the following relation,
\begin{equation}
4\pi R_0^2 \rho v^3\propto L,
\label{eq:kinetic_luminosity}
\end{equation}
is expected to hold for supernova ejecta with a sufficiently long-term energy supply. 
In other words, the kinetic luminosity of the flow is constant. 
This relation gives the density distribution immediately after the ejecta are affected by the energy injection. 
We denote the time when the ejecta following this relation are created by $t_0$. 
The ejecta are not freely expanding at this time. 
As the ejecta expand to the surrounding space, the density distribution would gradually be modified. 
Our goal is to derive the density distribution at the freely expanding stage. 

We regard the radial velocity $v$ as a Lagrangian coordinate and derive the density distribution as a function of $v$. 
We consider a concentric shell with inner and outer velocity coordinates $v$ and $v+\Delta v$. 
The inner and outer boundaries are located at $R=R_0(v)$ and $R=R_0(v+\Delta v)$ at $t_0$. 
The width $l_\mathrm{s}$ of the shell is initially given by,
\begin{equation}
l_\mathrm{s}(t_0)=R_0(v+\Delta v)-R_0(v)\simeq \frac{dR_0}{dv}\Delta v,
\label{eq:l_s}
\end{equation}
where the last expression is obtained for a sufficiently small $\Delta v$. 
The boundaries travel at the velocities $v$ and $v+\Delta v$ with time and reach $R=R_0(v)+v(t-t_0)$ and $R=R_0(v+\Delta v)+(v+\Delta v)(t-t_0)$ at $t$. 
For $t$ much longer than $t_0$, the initial radius $R_0(v)$ can be neglected and the inner and outer radii are given by $R\simeq vt$ and $R\simeq (v+\Delta v)t$.
The width $l_\mathrm{s}$ of the shell at $t$ can also be approximated as follows,  $l_\mathrm{s}(t)\simeq t\Delta v$.
The density of the shell at $t$ can be calculated by dividing the mass of the shell by the volume at $t$,
\begin{equation}
\rho(t,v)=\frac{4\pi R_0(v)^2l_\mathrm{s}(t_0)\rho(t_0,v)}{4\pi v^2t^3\Delta v}.
\end{equation}
Using the relation (\ref{eq:kinetic_luminosity}) and Equation (\ref{eq:l_s}), the density is written as follow,
\begin{equation}
\rho(t,v)\propto \frac{L}{4\pi v^5t^3}\frac{dR_0}{dv}.
\label{eq:rho_free}
\end{equation} 
The density is proportional to $t^{-3}$, reflecting the free expansion as expected. 
How the density depends on the velocity coordinate is determined by the factor $v^{-5}dR_0/dv$. 
The inverse of the latter term $dv/dR_0$ reflects the dependence of the velocity on the radial coordinate at $t_0$. 
When the velocity is simply proportional to the radius $R_0$, $v\propto R_0$, the derivative is a constant $dv/dR_0=\mathrm{Const.}$ and thus the density distribution is proportional to $v^{-5}$. 
On the other hand, when the velocity is a strongly growing function of the radius, e.g., $v\propto R_0^\lambda$ with $\lambda\gg 1$, the derivative $dv/dR_0\propto v^{(\lambda-1)/\lambda}$ is almost proportional to the velocity $v$, leading to $\rho \propto v^{-6}$. 
 Therefore, when the ejecta is powered by a constant energy injection and its kinetic luminosity is independent of the radial coordinate, the density structure of the ejecta at the free expansion stage is described by a power-law function of the velocity with an exponent between $-5$ and $-6$, depending on the radial velocity profile before entering the free expansion stage. 
Since the forward shock efficiently accelerates as it propagates in the outer ejecta, the shock velocity strongly grows with radius. 
Thus, density distributions close to the latter extreme case, $\rho \propto v^{-6}$, is expected to be realized rather than the former case, $\rho\propto v^{-5}$. 

However, the above consideration may be too idealistic. 
Although Equation (\ref{eq:kinetic_luminosity}) holds for the idealistic case, the energy transfer all the way to the outermost layer of the ejecta would not be so efficient. 
Thus, the kinetic luminosity can decrease with $R$ or $v$. 
From the results of our simulation, we found that the radial kinetic luminosity distribution slightly deviates from the uniform distribution and is close to $\propto R^{-1}$. 
Thus, the density profile in Equation (\ref{eq:rho_free}) should be slightly modified as follows,
\begin{equation}
\rho(t,v)\propto t^{-3}v^{-5}R_0^{-1}\frac{dR_0}{dv}.
\end{equation}
In this case, any power-law velocity profile, $v\propto R_0^\lambda$, leads to a power-law density distribution with an exponent $-6$. 
\begin{equation}
\rho(t,v)\propto t^{-3}v^{-6}. 
\end{equation}
This explains the reason why a simple power-law density distribution with an exponent $-6$ is realized in our simulation. 

In summary, power-law density distributions with exponents between $-5$ and $-6$ are expected to be realized in these cases, depending on the profiles of the kinetic luminosity and the radial velocity. 
Thus, the density slope significantly shallower than normal SNe is one of the important properties of SNe powered by a central engine lasting even after the breakout $t>t_\mathrm{br}$. 
When the energy injection is terminated before the breakout, $t<t_\mathrm{br}$, the subsequent dynamical evolution would be more similar to a point explosion, resulting in a steeper density slope. 

\subsection{One-zone Radiative Diffusion Model}\label{sec:one_zone}
The behaviour of the photospheric emission can also be obtained by applying the widely-used one-zone model for supernova light curves first derived by \cite{1980ApJ...237..541A} \citep[see, also,][]{1996snih.book.....A}. 
The luminosity of the emission evolves as
\begin{equation}
L_\mathrm{ph}(t)=\frac{4\pi f_\mathrm{L} R_\mathrm{ph,0}c}{3\kappa}\frac{E_\mathrm{th,0}}{M_\mathrm{ej}}\phi(t),
\label{eq:luminosity}
\end{equation}
where $f_\mathrm{L}$ ($\simeq 2$-$3$) is a numerical constant depending on the density structure of the ejecta and $R_\mathrm{ph,0}$ and $E_\mathrm{th,0}$ are the radius and the thermal energy of the ejecta when they are created at $t=t_0$. 
The dimensionless function $\phi(t)$ ($t>t_0$) introduced above governs the temporal evolution of the bolometric luminosity and is given as follows,
\begin{equation}
\phi(t)=\exp\left[-\frac{t-t_0}{t_\mathrm{d}}\left(1+\frac{t-t_0}{2t_\mathrm{h}}\right)\right],
\end{equation}
where $t_\mathrm{d}$ and $t_\mathrm{h}$ are the diffusion time of photons in the ejecta and the expansion timescale. 

From the calculations in the previous subsection, the ejecta turn out to be  well mixed with the hot bubble immediately after the breakout time $t=t_\mathrm{br}$ due to the acceleration of the forward shock in the outer ejecta, leading to the redistribution of the internal energy in the ejecta. 
Thus, we can roughly regard the breakout time $t=t_\mathrm{br}$ as the time of the creation of the ejecta having been powered by the energy injection at the centre, i.e., $t_0=t_\mathrm{br}$. 
In the following, we use the photospheric radius $R_\mathrm{ph,br}$ and the internal energy $E_\mathrm{int,br}$ at the breakout as $R_\mathrm{ph,0}$ and $E_\mathrm{th,0}$, both of which can easily be estimated from the density profile of the supernova ejecta and the semi-analytic model in the previous section. 
Even after the breakout, the internal energy gradually increases with time until the energy injection is terminated (see Figure \ref{fig:evolution}). 
However, the maximum value of the internal energy only differs from that at the breakout by a factor of a few. 
The increase in the internal energy is due to the dissipation of the kinetic energy of the relativistic wind. 
Since the gas leaking from the hot bubble can directly transport its energy outside the photosphere, it is unclear whether the dissipated energy really contributes to the increase in the radiation energy after the breakout. 
Therefore, we simply use the photospheric radius and the internal energy at the breakout. 
The photospheric radius $R_\mathrm{ph,br}$ can be calculated from the optical depth, Equation (\ref{eq:tau}), for the density profile at time $t=t_\mathrm{br}$, Equation (\ref{eq:ejecta_density}). 
Since the photosphere is still located in the unshocked outer ejecta, we obtain
\begin{equation}
R_\mathrm{ph,br}=w_\mathrm{c}v_\mathrm{ej}t_\mathrm{br}
\left[\frac{4\pi(w_\mathrm{c}v_\mathrm{ej}t_\mathrm{br})^2}{(n-1)\kappa f_3M_\mathrm{ej}}+w_\mathrm{c}^{n-1}\right]^{\frac{1}{1-n}}.
\end{equation}
Furthermore, Equation (\ref{eq:E_th}) gives the thermal energy at the breakout, 
\begin{equation}
E_\mathrm{int,br}=\frac{2-\gamma}{1+3\alpha(\gamma-1)}f_\mathrm{br}E_\mathrm{sn}.
\end{equation}
The radius and the thermal energy are found to be $R_\mathrm{ph,br}=4.0\times 10^{14}$ cm and $E_\mathrm{int,br}=1.5\times 10^{51}$ erg for the parameters adopted in our simulation. 
Thus, the luminosity at $t=t_0$ yields,
\begin{eqnarray}
L_\mathrm{ph}(t_0)&=&3.8\times 10^{43}f_\mathrm{L}\ \mathrm{erg}\ \mathrm{s}^{-1}
\nonumber\\
&&\times
\left(\frac{\kappa}{0.1\ \mathrm{cm^{2}\ g}^{-1}}\right)^{-1}
\left(\frac{R_\mathrm{ph,br}}{4\times 10^{14}\ \mathrm{cm}}\right)
\nonumber\\
&&\times
\left(\frac{M_\mathrm{ej}}{10\ M_\odot}\right)^{-1}
\left(\frac{E_\mathrm{sn}}{10^{51}\ \mathrm{erg}}\right).
\end{eqnarray}
Taking $f_\mathrm{L}=2$-$3$ \citep{1980ApJ...237..541A}, we obtain a luminosity of the order of $10^{44}\ \mathrm{erg}\ \mathrm{s}^{-1}$, which again agrees with bolometric luminosities of SLSNe-I. 

Finally, we calculate the total radiated energy $E_\mathrm{rad}$. 
We simply estimate the timescales of the emission as follows,
\begin{equation}
t_\mathrm{d}=\frac{\kappa M_\mathrm{ej}}{cR_\mathrm{ph,br}},
\end{equation}
and
\begin{equation}
t_\mathrm{h}=\frac{R_\mathrm{ph,br}}{v_\mathrm{c}},
\end{equation}
where 
\begin{equation}
v_\mathrm{c}=\sqrt{\frac{2f_\mathrm{br}E_\mathrm{sn}}{M_\mathrm{ej}}},
\end{equation}
is the average velocity of the ejecta after an energy of $f_\mathrm{br}E_\mathrm{sn}$ has been deposited. 
The integration of Equation (\ref{eq:luminosity}) with respect to time $t$ from $0$ to $\infty$ leads to the total radiated energy $E_\mathrm{rad}$. 
The integration can be approximated as follows,
\begin{equation}
E_\mathrm{rad}=\int_0^\infty L_\mathrm{ph}(t)dt
\simeq \sqrt{t_\mathrm{d}t_\mathrm{h}}L_\mathrm{ph}(t_0).
\end{equation}
Thus, emission at a luminosity of $L_\mathrm{ph}(t_0)$ lasts for a timescale of 
\begin{eqnarray}
\sqrt{t_\mathrm{d}t_\mathrm{h}}&=&
2^{-1/4}c^{-1/2}\kappa^{1/2}f_\mathrm{br}^{-1/4}M_\mathrm{ej}^{3/4}E_\mathrm{sn}^{-1/4}
\nonumber\\
&=&
9.6\times 10^{6}\ \mathrm{s}
\left(\frac{\kappa}{0.1\ \mathrm{cm^{2}\ g}^{-1}}\right)^{1/2}
\nonumber\\
&&\times
\left(\frac{M_\mathrm{ej}}{10\ M_\odot}\right)^{3/4}
\left(\frac{E_\mathrm{sn}}{10^{51}\ \mathrm{erg}}\right)^{-1/4}.
\label{eq:t_emission}
\end{eqnarray}
The total radiated energy yields
\begin{eqnarray}
E_\mathrm{rad}&=&3.7\times10^{50}f_\mathrm{L}\ \mathrm{erg}
\nonumber\\
&&\times
\left(\frac{\kappa}{0.1\ \mathrm{cm^{2}\ g}^{-1}}\right)^{-1/2}
\left(\frac{R_\mathrm{ph,br}}{4\times 10^{14}\ \mathrm{cm}}\right)
\nonumber\\
&&\times
\left(\frac{M_\mathrm{ej}}{10\ M_\odot}\right)^{-1/4}
\left(\frac{E_\mathrm{sn}}{10^{51}\ \mathrm{erg}}\right)^{3/4}.
\end{eqnarray}
For the parameters adopted here, the duration and the total radiated energy of the emission are estimated to be $(t_\mathrm{d}t_\mathrm{h})^{1/2}\simeq 1.1\times 10^2$ days and $E_\mathrm{rad}\simeq 10^{51}$ erg. 
Therefore, a significant fraction of the internal energy is expected to be released as photons before being lost via adiabatic expansion.

\begin{figure}[tbp]
\begin{center}
\includegraphics[scale=0.5,bb=0 0 453 680]{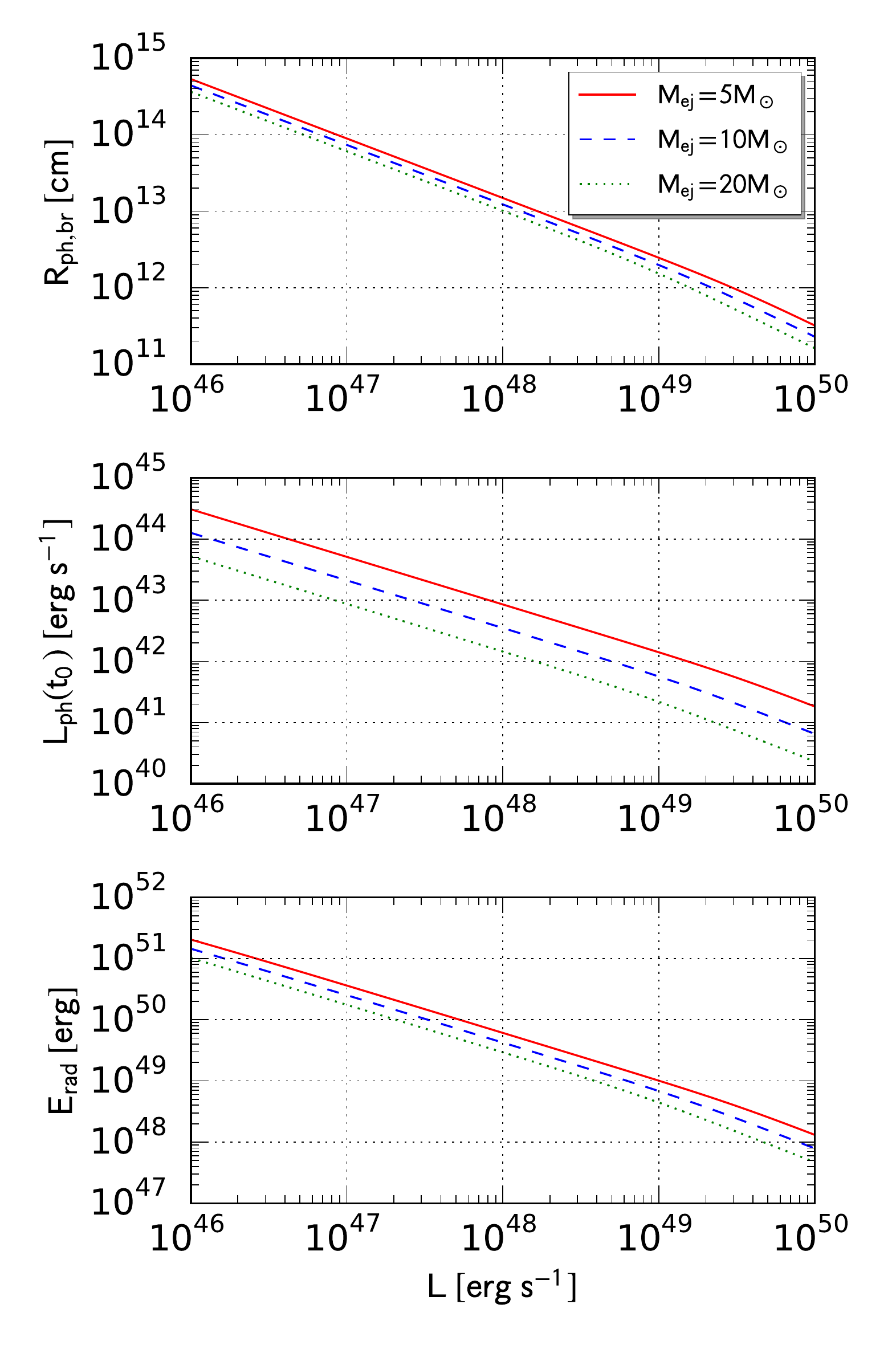}
\caption{Photospheric radius $R_\mathrm{ph,br}$ (top panel), the luminosity $L_\mathrm{ph}(t)$ at $t=t_0$ (middle panel), and the total radiated energy $E_\mathrm{rad}$ (bottom panel) as functions of the energy injection rate $L$. 
The solid, dashed, and dotted lines in each panel correspond to models with ejecta masses of $M_\mathrm{ej}=5,$ $10$, and $20\ M_\odot$. 
The other free parameters characterizing the ejecta are fixed to be $m=1$, $n=10$, $w_\mathrm{c}=0.1$, and $E_\mathrm{sn}=10^{51}\ \mathrm{erg}$. }
\label{fig:emission}
\end{center}
\end{figure}

Figure \ref{fig:emission} shows the photospheric radius $R_\mathrm{ph,br}$, the luminosity $L_\mathrm{ph}$ at $t=t_0$, and the total radiated energy $E_\mathrm{rad}$ as functions of the energy injection rate $L$ for different ejecta masses, $M_\mathrm{ej}=5,$ $10$, and $20\ M_\odot$. 
For smaller injection rates, the breakout occurs at larger radii and the emission becomes brighter, leading to an efficient conversion of the injected energy to radiation energy. 
For higher injection rates, $L\sim 10^{48-50}\ \mathrm{erg}\ \mathrm{s}^{-1}$, on the other hand, the total radiated energy is only a small fraction of the injected energy because the ejecta significantly suffer from adiabatic loss.

\subsection{Supernovae Powered by Central Engine}
Our results indicate that the additional energy injection into supernova ejecta could result in two different types of transients. 

\subsubsection{Superluminous Supernovae}
As shown in Figure \ref{fig:emission}, lower energy injection rates lead to high radiation efficiencies. 
The timescale of the release of the thermal energy in the ejecta via radiative diffusion and the luminosity of the emission are consistent with those of SLSNe-I. 
In addition, different layers of the supernova ejecta, which are supposed to be well stratified initially, are mixed after the breakout of the hot bubble. 
The spectral properties of the emission from the ejecta should significantly be influenced by the mixing. 
Since chemical elements are allowed to spread throughout layers with various radial velocities, the spectra would be subject to severe line blending and become featureless, which could also explain featureless spectra of SLSNe-I. 

\cite{2016MNRAS.458.3455M} investigated the spectral formation in SLSNe-I by combining a simple ejecta model with radiative transfer calculations. 
They assumed that the density structure of the ejecta was represented by a power-law function of the radial coordinate with an exponent $-7$ and found that their calculations well reproduced the spectra of several SLSNe-I around the maximum. 
Our simulation shows the radial density profile characterized by a power-law function of the radial coordinate with an exponent $-6$ (see, Figure \ref{fig:radial_final}) and we have given a theoretical explanation for the density structure in Section \ref{sec:density_structure}. 
Our density profile is slightly shallower, $\rho \propto R^{-6}$, rather than $\rho\propto R^{-7}$. 
Since \cite{2016MNRAS.458.3455M} did not show results with different exponents, it is unclear whether the power-law density distribution realized in our simulation could reproduce spectra of SLSNe-I. 
Nevertheless, our finding that single power-law density distributions are realized in supernova ejecta with an central energy injection can be used to better constrain models responsible for SLSNe-I. 

\subsubsection{Broad-lined Ic Supernovae}
On the other hand, models with high energy injection rates can convert only a small fraction of the injected energy into radiation. 
This is mainly due to adiabatic loss. 
Ejecta with higher energy injection rates experience the breakout of the hot bubble at early epochs. 
The breakout time $t_\mathrm{br}=f_\mathrm{br}t_\mathrm{c}$ is much smaller than the timescale of radiative diffusion, $(t_\mathrm{d}t_\mathrm{h})^{1/2}$, which means that most of the injected energy is used for the expansion of the gas before being released as thermal radiation. 
Therefore, supernova ejecta with a high energy injection rate consume the additional energy to increase its kinetic energy instead of outshining brightly. 
As a consequence, these ejecta are observed as supernovae with kinetic energies of the order of $10^{52}$ erg. 
Furthermore, we can also expect featureless spectra with significant line blending. 
These properties agree with those of broad-lined Ic SNe, which are closely linked with GRBs. 

Recent radio observations have found broad-lined Ic SNe whose properties are similar to SNe associated with GRBs but without gamma-ray detection or bright X-ray emission, such as SN 2009bb \citep{2010Natur.463..513S,2010ApJ...725....4B,2011ApJ...728...14P} and 2012ap \citep{2014ApJ...797..107M,2015ApJ...799...51M,2015ApJ...805..187C}.  
These SNe are characterized by bright radio emission, which is interpreted as synchrotron emission from blast waves at velocities of $0.7c$-$0.9c$. 
Optical and radio observations of these events indicate that a considerable amount of kinetic energy, $10^{49-50}$ erg, is coupled to the relativistic component of the ejecta, suggesting relatively flat kinetic energy distributions. 
The presence of trans-relativistic ejecta led the authors to conclude that they are SNe powered by some central engine activity. 
In addition, these events have a lot in common with an under-luminous class of GRBs (low-luminosity GRBs), such as GRB 980425/SN 1998bw \citep{1998Natur.395..663K,1998Natur.395..670G}, GRB 060218/SN 2006aj \citep{2006Natur.442.1008C,2006Natur.442.1011P,2006Natur.442.1014S,2006Natur.442.1018M}, and GRB 100316D/SN 2010bh \citep{2011MNRAS.411.2792S}, with respect to optical and radio properties.  
These radio-loud SNe and low-luminosity GRBs exhibit no signature of harbouring ultra-relativistic jets, suggesting that the additional energy injection may be realized in a quasi-spherical way rather than highly collimated jet injection. 

The flat kinetic energy distribution obtained by our simulation (see Figure \ref{fig:spec}) agrees with those of radio-loud SNe and low-luminosity GRBs. 
Our simulation only deals with an energy injection at a rate of $L=10^{46}\ \mathrm{erg}\ \mathrm{s}^{-1}$. 
However, since the hydrodynamic evolution can be scaled by a single timescale $t_\mathrm{c}$, similar kinetic energy distributions must be realized even when higher energy injection rates $L\sim 10^{49-50}\ \mathrm{erg}\ \mathrm{s}^{-1}$ are assumed. 
In fact, radio observations of SN 2012ap suggest that the radio-emitting ejecta is characterized by a power-law density profile with an exponent $-6$ \citep{2015ApJ...805..187C}, which agrees with the density structure of the ejecta realized in our simulation. 
This remarkable agreement supports the possibility that quasi-spherical energy injection into supernova ejecta at high injection rates can produce broad-lined Ic SNe with trans-relativistic ejecta. 

\subsubsection{Possible Link between SLSNe-I and GRBs}
In this scenario, similar energy injection mechanisms are responsible for both SLSNe-I and broad-lined Ic SNe. 
Since some broad-lined Ic SNe are associated with GRBs, this also suggests a possible link between SLSNe-I and GRBs. 
This has already been pointed out by, e.g., \cite{2015MNRAS.454.3311M}. 
Very recently, spectroscopic observations of two SLSNe-I, LSQ14an and SN 2015bn in their nebular phases by \cite{2016arXiv160802994J} and \cite{2016ApJ...828L..18N} revealed optical spectra reminiscent of those of broad-lined Ic SNe, such as SN 1998bw, further supporting the link between SLSNe-I and GRBs. 

In addition, several follow-up observations of host galaxies of SLSNe-I and GRBs suggest that they have a lot in common, such as their preference for environments with low metallicity and high specific star formation rates \citep{2011ApJ...727...15N,2014ApJ...787..138L,2015MNRAS.449..917L,2016arXiv160504925C}. 
The scenario described above can naturally explain the similarities in explosion sites of SLSNe-I and GRBs. 
Further observational studies would shed light on the progenitor system and the physical conditions producing SLSNe-I and GRBs.

\subsection{Implications for Properties of SLSNe}
\subsubsection{Power Source of Bright Emission and Early Light Curve}
In this study, we focus on the internal energy stored in the ejecta until the breakout of the forward shock from the photosphere located in the outer ejecta and regard it as the potential power source of the bright emission from SLSNe-I. 
The energy injection is suddenly terminated at $t=10.1t_\mathrm{c}$. 
We then estimate the luminosity and the total radiated energy of the emission by using the Arnett's model treating the diffusion of photons throughout the ejecta under the one-zone approximation \citep{1980ApJ...237..541A}. 
On the other hand, several studies provide more specific emission models. 
\cite{2016ApJ...821...36K} investigated supernova ejecta with an energy injection from a magnetar at the centre of the ejecta. 
Their 1D radiation-hydrodynamic simulations showed the formation of a geometrically thin shell. 
Based on the results of their simulations and an analytic model, they calculated the temporal evolution of the emission powered by the energy injection. 
In their analytic considerations, they separately treat the internal energy deposited by the forward shock in the outer ejecta and that stored in the whole ejecta, which gradually diffuses out through the ejecta. 
This scenario can explain SLSNe exhibiting double-peaked light curves, such as SN 2006oz \citep{2012A&A...541A.129L}, LSQ14bdq \citep{2015ApJ...807L..18N,2016MNRAS.457L..79N}, and DES14X3taz \citep{2016ApJ...818L...8S}. 
The first peak is attributed to the former energy source, the internal energy deposited by the forward shock into the outer ejecta, while the second peak is explained by the usual magnetar-powered emission whose energy source is the internal energy deposited by the magnetar spin-down into the ejecta. 
In their model, the former energy source only contributes to a small fraction of the total radiated energy. 

On the other hand, our calculations in Section \ref{sec:one_zone} do not separately treat the internal energies in the inner outer parts of the ejecta and assume that the internal energy stored in the whole ejecta is released as radiation within the emission timescale given by Equation (\ref{eq:t_emission}). 
We only consider the internal energy supplied by the central engine until the breakout and ignore the tail of the energy injection rate expected in the magnetar spin-down scenario, which may underestimate the internal energy available for the bright emission. 
Although the continuously injected energy even after the breakout can contribute to the bright emission, our calculations suggest that the internal energy accumulated in the ejecta before the breakout can account for the total radiated energy of $10^{51}$ erg for cases with low energy injection rates, $L\sim 10^{46}$ erg s$^{-1}$. 

Our numerical simulation also demonstrates that the breakout of materials powered by the energy injection is even more drastic. 
The forward shock propagating in the outer ejecta reaches sub-relativistic velocities before emerging from the photosphere in the ejecta. 
The injected energy is directly carried by the leaking hot gas toward outermost layers. 
This leads to a remarkable contrast to the model by \cite{2016ApJ...821...36K}, where the early emission from SLSNe is realized by photons diffusing out of the well-stratified layers of spherical ejecta. 
The situation realized in our simulation is similar to the supernova shock breakout in a collapsing massive star, although the shape of the forward shock front significantly deviates from spherical symmetry. 
Whether the ejecta are gradually heated by photons from the shell or spontaneously heated by the forward shock would affect the properties of the early emission from the ejecta, such as rising times, cooling rates, and so on. 
Recent photometric and spectroscopic observations of the SLSN-I Gaia16apd \citep{2016arXiv161106993N,2016arXiv161110207K} found by the {\it Gaia} survey showed prominent UV emission in the earliest phase of the evolution. 
Our results suggest that this UV emission may be attributed to the breakout of the forward shock from the photosphere in the supernova ejecta. 
As shown in Figure \ref{fig:photo}, the isotropic luminosity and the radiation temperature immediately after the breakout, $t=6.0t_\mathrm{c}$, reach a few $10^{44}$ erg s$^{-1}$ and $\sim 10^5$ K, which suggests bright emission with an extremely blue spectrum, although more sophisticated treatments of radiative transport are required. 
In addition, the initial phase of the supernova shock breakout emission is sensitive to the geometry of the forward shock emerging the photosphere and the viewing angle \citep{2010ApJ...717L.154S,2016ApJ...825...92S}. 
Thus, early light curves and spectra of SNe driven by central engines could provide important information on the geometry of the forward shock propagating in the outer ejecta.

\subsubsection{Late Time Behaviors}
As shown in Figure \ref{fig:radial_final}, the density distribution after the termination of the energy injection is well represented by a single power-law function of the radial or the velocity coordinate. 
Simple power-law density distributions predict continuous changes in the properties of the photospheric emission, such as the photospheric velocity, the colour temperature, and so on, as long as the photosphere is still located in the power-law part. 
However, we should note that a late-time energy injection may modify the density structure. 
Although we suddenly stop injecting energy into the ejecta, a gradually terminated energy injection, which is expected in magnetar spin-down scenario, may result in different distributions. 

Recent spectroscopic observations of two SLSNe-I in the nebular phase \citep{2016arXiv160802994J,2016ApJ...828L..18N} revealed that the nebular spectra were dominated by prominent emission lines. 
Therefore, the ejecta should be exposed to high-energy electrons and/or photons so that the gas is kept ionized even at the nebular stage. 
In normal SNe, radioactive decays of $^{56}$Ni and $^{56}$Co provide the ejecta with non-thermal electrons and gamma-ray photons. 
However, SLSNe may not be powered by the radioactivity. 
The energy injection from the central engine would be realized in the form of relativistic particles and high-energy photons in the same way as pulsar winds. 
The late-time activity of the central engine and the density structure of the ejecta are therefore of significant importance in explaining the ejecta in the nebular stage and better constraining existing model for SLSNe-I. 
 
\subsubsection{High-energy Radiation Signatures}
The breakout of the forward shock from the supernova ejecta is probabily followed by the escape of high-energy photons and particles into the interstellar space, which may potentially distinguish central engine scenarios from other scenarios for SLSNe-I. 
\cite{2014MNRAS.437..703M} considered the ionization structure of spherical supernova ejecta illuminated by the powerful emission from a magnetized millisecond pulsar. 
Electrons and positrons created and accelerated to high energies in the vicinity of the neutron star could produce high-energy photons via pair cascades, which can ionize the ejecta. 
The ionization structure of the ejecta is determined by the balance between ionization by high-energy photons and particles and recombination of free electrons with ions. 
As the ejecta expand, the decreasing ejecta density makes recombination less efficient and thus the ionization front propagates outward in the ejecta. 
They argued that the breakout of the ionization front from the ejecta could give rise to bright X-ray or UV emission. 

The multi-dimensional ejecta structure due to the development of the Rayleigh-Taylor instability influences the efficiency of the escape of high-energy particles from the ejecta. 
As we have shown in Figure \ref{fig:snap2} and discussed in Section \ref{sec:hot_bubble}, materials leaking from the hot bubble accelerate outside the shell and their Lorentz factors can be well above unity depending on the injection condition of the relativistic wind. 
This ``shredding'' of the ejecta by the hydrodynamic instability has already been pointed out by \cite{2003ApJ...589..871A}, who studied ultrahigh-energy cosmic-rays produced by magnetar activities and their leakage from the surrounding supernova ejecta.

The multi-dimensional structure with low-density channels tends to help high-energy particles escape easily from the surrounding medium because these particles can stream through the channels and propagate farther than the case of well-stratified ejecta. 
In other words, the mean free path of high-energy particles can effectively be enhanced due to the patchy density structure. 
However, it is important to note that channels in the ejecta are curved and twisted and thus a particle travelling along a straight trajectory in a channel easily hits the wall of the channel. 
This means that the shredded ejecta can enhance the mean free path of the species of interest only up to the physical scale of the channel. 
Once the mean free path of the species for an averaged density becomes much longer than the physical scale of the channels, the shredded ejecta are unlikely to greatly affect the escape fraction of high-energy particles. 
Therefore, how efficiently the multi-dimensional effect can enhance the transparency of the ejecta to high-energy particles significantly depends on the physical scale of the channels. 

The linear analysis of the Rayleigh-Taylor instability predicts that perturbations with shorter wavelengths grow faster \citep[e.g.][]{1961hhs..book.....C}. 
In supernova ejecta, several effects of radiative transport, such as the radiative diffusion length, determine the shortest unstable wavelength \citep{1978ApJ...219..994C}. 
Since our simulation does not treat radiative transport, the minimum physical scale must be set by the resolution of the simulation. 
Channels with smaller physical scales can be realized in simulations with higher resolutions. 
Furthermore, future 3D simulations may reveal ejecta structure with different morphology and channels with different size distributions. 
Therefore, the minimum physical scale determined by the radiative transport effects and the comparison with the mean free path of high-energy photons, electrons, positrons, and ions should be studied in detail to quantitatively determine the escape fraction of these particles and the ionization states of different layers of the ejecta. 

We may expect the possibility that these highly relativistic flows are predominantly composed of cold leptons and baryons and they dissipate their kinetic energies through shocks outside the photosphere, leading to flare activities associated with the dissipation and characterized by high-energy emission with non-thermal spectra. 
Since the flows are driven by the energy injection from the relativistic wind, the energy flux of each relativistic flow is basically determined by that of the relativistic wind. 
Thus, the isotropic luminosity of the high energy emission would be of the order of the energy injection rate at the centre, $L$.

\subsection{Implications for Magnetar Spin-down Scenario}
In this study, we simply inject energy into the supernova ejecta at a constant rate and do not assume any specific mechanism responsible for the energy injection. 
In the following, we briefly mention implications for the magnetar spin-down scenario.

We consider a neutron star with typical values of the radius $R_\mathrm{ns}=10$ km and the moment of inertia $I_\mathrm{ns}\sim 10^{45}$ g cm$^{2}$. 
The rotational energy of the neutron star is given by $E_\mathrm{rot}=I_\mathrm{ns}\Omega_\mathrm{i}^2/2$, where $\Omega_\mathrm{i}$ is the initial frequency. 
Therefore, in order for the neutron star to deposit a total energy of the order of $10^{52}$ erg, it should be rotating at an initial frequency of $\Omega_\mathrm{i}\sim 4\times 10^3$ $\mathrm{s}^{-1}$, corresponding to an initial period $P_\mathrm{i}=2\pi/\Omega_\mathrm{i}$ of the order of $1$ ms. 

In the magnetar scenario, the rotational energy is lost via magnetic dipole radiation. 
For a given dipole magnetic field strength $B$, the neutron star loses its rotational energy at a spin-down rate of
$L\simeq E_\mathrm{rot}/t_\mathrm{ch}$ \citep[e.g.][]{1983bhwd.book.....S} at $t<t_\mathrm{ch}$. 
The timescale $t_\mathrm{ch}$ characterizing the energy loss is given by
\begin{equation}
t_\mathrm{ch}=\frac{6I_\mathrm{ns}c^3}{B^2R_\mathrm{ns}^6\Omega_\mathrm{i}^2}
=4.1\times10^3
I_\mathrm{ns,45}B_{15}^2R_\mathrm{ns,6}^6P_\mathrm{i,-3}^{2}
\ \mathrm{s}.
\end{equation}
The physical quantities are expressed by $Q=10^nQ_n$ in cgs units. 
Until the characteristic time, $t<t_\mathrm{ch}$, the spin down of the neutron star deposits the rotational energy at a rate,
\begin{equation}
L\simeq \frac{B^2R_\mathrm{ns}^6\Omega_\mathrm{i}^4}{6c^3}\sim
10^{49}
B_{15}^2R_\mathrm{ns,6}^6P_\mathrm{i,-3}^{-4}\ \mathrm{erg\ s}^{-1}.
\end{equation}

Therefore, strong magnetic field strengths of the order of $B=10^{15}$ G, which are typical for Galactic magnetars, yield energy injection timescales of $10^3$ s. 
In our model, energy injections with such high rates lead to radiatively inefficient explosions, which are supposed to produce transients like broad-lined Ic SNe. 
On the other hand, in order to produce radiatively efficient explosions or SLSNe-like transients, the magnetic field strength should be $10^{13}$-$10^{14}$ G. 
The same conclusion had been reached by \cite{2015MNRAS.454.3311M}, who considered the magnetar scenario by a one-zone model with energy supplies from the magnetar spin-down and the radioactive decay of $^{56}$Ni and energy losses via adiabatic cooling and radiative diffusion. 

\subsection{Other Remarks}
Finally, we describe some remarks and future prospects. 

First, the presence of the symmetry axis in our simulation would affect the dynamical evolution of the ejecta. 
As shown in Figures \ref{fig:snap1} and \ref{fig:snap2}, the deviation from the spherical symmetry is most prominent around the symmetry axis at $r=0$. 
At the axis, collisions of incoming and reflected flows produce large perturbations, from which the Rayleigh-Taylor instability efficiently develops. 
As a result, the forward shock emerges from the supernova ejecta in a bipolar fashion. 
This is inevitable as long as we perform two-dimensional simulations. 
In three-dimensional simulations without any preferred direction, deviations from spherical symmetry would equally grow along all radial directions. 
In addition, the growth of the Rayleigh-Taylor instability may differ in the presence of magnetic fields \citep[e.g.][]{2007ApJ...671.1726S}. 
In the context of pulsar wind nebulae, multi-dimensional numerical modellings of the wind-ejecta interaction have been attempted \citep[e.g.][]{2003MNRAS.344L..93K,2004A&A...421.1063D}. 
\cite{2014MNRAS.438..278P} performed both two- and three-dimensional relativistic magnetohydrodynamics simulations. 
They clearly demonstrated the non-linear development of the magnetic Rayleigh-Taylor instability. 
However, it is difficult to compare the power spectra of the magnetic field in the three-dimensional simulation with those of two-dimensional counterparts because of the limited resolution. 
Thus, further sophisticated numerical investigations are strongly demanded. 
We should also consider the possibility that bipolar structure is realized in reality. 
In the magnetar scenario, the energy injection is due to the magnetic dipole radiation, which is anisotropic in nature. 
Recent spectroscopic observations of SLSN-I 2015bn by \cite{2016arXiv160702353I} reported the detection of significant polarisation at both pre- and post-maximum stages. 
They pointed out the possibility that the energy injection is realized in a similar manner to broad-lined Ic SNe associated with GRBs. 

Next, the coupling between gas and radiation should be treated in a more sophisticated way. 
At early stages of the dynamical evolution of the ejecta, the shock wave propagates in sufficiently dense media. 
Thus, the assumption of the strong coupling between gas and radiation is appropriate. 
After the breakout of the forward shock, however, photons having been coupled to gas can be efficiently advected to outside the photosphere by the flows driven by the leaking hot bubble. 
Then, the radiation pressure no longer pushes the outermost layer of the ejecta. 
This effect is not treated in our simulation employing an ideal gas law with an index $4/3$. 
Radiatively cooling shock waves are known to be subject to a number of instabilities \citep{2016MNRAS.459.2188B}. 
Furthermore, as we have noted in the previous section, the breakout emission from the forward shock emerging from the photosphere highly depends on the geometry of the forward shock. 
Multi-dimensional radiation-hydrodynamic simulations would resolve these problems. 

Roles of magnetic fields should also be investigated in detail. 
In central engine scenarios, the presence of strong magnetic fields is essential in extracting the rotational energy of the compact remnant embedded in supernova ejecta. 
Specifically, in the magnetar scenario, the spin down of a magnetar via magnetic braking determines the energy deposition rate. 
Furthermore, magnetic fields may also play key roles in determining whether the energy deposition is realized in a quasi-spherical or a highly aspherical way, as demonstrated by \cite{2008MNRAS.383L..25B,2009MNRAS.396.2038B}. 

As we have demonstrated by our simulation, the process redistributing the energy injected at the centre throughout the supernova ejecta is multi-dimensional by nature. 
This strongly indicates that further sophisticated numerical investigations on supernova explosions with central engine activities are required for the ultimate understanding of extraordinary SNe.

\section{SUMMARY}\label{sec:summary}
In this paper, we consider the hydrodynamical evolution of supernova ejecta with energy injection from a central compact object. 
The supernova ejecta considered in this study have the inner and outer parts with shallow and steep density gradients. 
The energy injection is realized as a relativistic wind. 
The collision between the wind and the ejecta creates a couple of shock waves, the forward and reverse shocks, which efficiently convert the kinetic energy of the relativistic wind into the internal energy of the shocked gas. 
The dissipation of the kinetic energy leads to the formation of a hot bubble pushing a geometrically thin shell. 
Early stages of the dynamical evolution of the shell can be described by a self-similar solution. 
At later stages, on the other hand, the shell is subject to an efficient mixing due to the development of the Rayleigh-Taylor instability and thus cannot be treated analytically. 
We have performed a numerical simulation of its hydrodynamical evolution by using our special relativistic hydrodynamic code equipped with AMR. 
Our simulation follows the non-linear development of the instability, through which the hot bubble completely destroys the geometrically thin shell. 
As a result, the hot bubble and the supernova ejecta are well mixed. 
This process occurs immediately after the forward shock reaches the interface between the inner and outer ejecta and the moment of its onset can be predicted by using the self-similar solution. 
The efficient mixing of the injected gas and the ejecta and the subsequent redistribution of the injected energy are not seen in studies assuming spherical symmetry and make the dynamical evolution of the ejecta qualitatively different from the picture of the spherical shell stacking in a deep layer of the ejecta. 
Therefore, the multi-dimensional effects on the properties of the supernova ejecta and their potential to give rise to bright emission should be thoroughly examined. 

The material leaking from the hot bubble efficiently redistributes the energy injected at the centre throughout the whole ejecta. 
Using spatial distributions of physical variables obtained by our simulation, we calculate the distributions of the mass, the kinetic, and the internal energy of the ejecta after an energy 10 times larger than the initial kinetic energy of the supernova ejecta has been injected. 
Furthermore, we investigate properties of the expected emission from the ejecta by identifying the photosphere in the ejecta. 
The luminosity and temperature of the photospheric emission are estimated to be of the order of $10^{44}\ \mathrm{erg}\ \mathrm{s}^{-1}$ and $10^4$ K, which could explain the high brightness and relatively blue spectra of SLSNe-I. 
We apply the radiative diffusion model presented by \cite{1980ApJ...237..541A} to estimate the energy lost from the ejecta as radiation. 
When an energy injection rate of $L=10^{46}$ erg is adopted, the total radiated energy is found to be $10^{51}$ erg, which also agrees with those of SLSNe-I. 
In addition, outflows travelling at relativistic speeds may appear after the breakdown of the shell as shown in Figure \ref{fig:snap2}. 
These highly relativistic outflows create low-density channels in the ejecta, through which high-energy particles possibly escape into the interstellar space. 
The resultant shredded ejecta are a key to unveiling the escape fraction of high-energy photons, electrons, positions, and ions probably produced in the vicinity of the central engine and the ionization states of the ejecta. 

\acknowledgments
We appreciate the anonymous referee for his or her constructive comments, which helped us to improve the manuscript. 
Numerical calculations were in part carried out on the Cray XC30 system at Center for Computational Astrophysics, National Astronomical Observatory of Japan. 
AS is supported by Grant-in-Aid for JSPS Research Fellow (26$\cdot$10618). 
The work by KM is partly supported by Japan Society for the Promotion of Science (JSPS) KAKENHI Grant 26800100 and by World Premier International Research Center Initiative (WPI Initiative), MEXT, Japan.

\appendix
\section{DERIVATION OF THE SELF-SIMILAR SOLUTION}\label{sec:self_similar}
In this section, we derive the self-similar solution describing the propagation of a blast wave in freely expanding ejecta with a power-law density profile. 
The dynamical evolution of a spherical blast waves propagating in a medium in self-similar ways is a classical but important problem in astrophysics and has a long histroy. 
Thus, there are a lot of studies focusing on this problem  \citep[e.g.][]{sedov46,vonNeumann47,1950RSPSA.201..159T,1982ApJ...258..790C,1993PhFl....5.1035W,1998ApJ...499..282J,2002A&A...395.1023D,2016PASJ...68...22M} and one one can find how to solve such problems in some textbooks or review papers \citep{1959sdmm.book.....S,1966egct.book.....Z,1988RvMP...60....1O,1996sssi.book.....B}.
\subsection{Shock Jump Conditions}
The position of the shock front in the supernova ejecta is denoted by $R_\mathrm{s}$. 
The shock radius is assumed to be a power-law function of time $t$ with an exponent $\alpha$,
\begin{equation}
R_\mathrm{s}=At^{\alpha}.
\end{equation}
The shock velocity is given by differentiating the shock radius with respect to time $t$,
\begin{equation}
V_\mathrm{s}=\frac{dR_\mathrm{s}}{dt}=\alpha \frac{R_\mathrm{s}}{t}.
\end{equation}
The shock jump conditions for physical variables are given by the flux conservation in the rest frame of the shock. 
We denote the post-shock values of the density, the velocity, and the pressure by $\rho_\mathrm{f}$, $u_\mathrm{f}$, and $p_\mathrm{f}$. 
From the jump conditions, we can express the post-shock values in terms of the shock velocity $V_\mathrm{s}$ and the pre-shock values $\rho_\mathrm{ej}$, $u_\mathrm{ej}$, and $p_\mathrm{ej}$ as follows,
\begin{equation}
u_\mathrm{f}=\frac{2V_\mathrm{s}+(\gamma-1)u_\mathrm{ej}}{\gamma+1},\ \ \ 
\rho_\mathrm{f}=\frac{\gamma+1}{\gamma-1}\rho_\mathrm{ej},\ \ \ 
p_\mathrm{f}=\frac{2}{\gamma+1}\rho_\mathrm{ej}(V_\mathrm{s}-u_\mathrm{ej})^2.
\label{eq:post_shock1}
\end{equation}
Furthermore, the sound speed $c_\mathrm{s,f}$ at the shock front is given by,
\begin{equation}
c_\mathrm{s,f}=\sqrt{\frac{\gamma p_\mathrm{f}}{\rho_\mathrm{f}}}
=\frac{\sqrt{2\gamma(\gamma-1)}}{\gamma+1}(V_\mathrm{s}-u_\mathrm{ej}).
\label{eq:post_shock2}
\end{equation}

While the shock wave is propagating in the inner ejecta, whose density profile is given in Equation (\ref{eq:ejecta_density}), the pre-shock density at the shock front should evolve as,
\begin{equation}
\rho_\mathrm{ej}=Dt^{m-3}R_\mathrm{s}^{-m}=A^{-m}Dt^{\delta},
\end{equation}
with
\begin{equation}
D=\frac{f_\mathrm{3}M_\mathrm{ej}}{4\pi}(\omega_\mathrm{c}v_\mathrm{ej})^{m-3},
\end{equation}
and
\begin{equation}
\delta=-\alpha m+m-3.
\end{equation}
Since the ejecta are freely expanding, the pre-shock velocity is given by
\begin{equation}
u_\mathrm{ej}=\frac{R_\mathrm{s}}{t}=At^{\alpha-1}.
\end{equation}
Therefore, the post-shock values of the velocity, the sound speed, and the density are expressed as follows,
\begin{equation}
u_\mathrm{f}=\frac{2\alpha+\gamma-1}{\gamma+1}At^{\alpha-1},\ \ \ 
c_\mathrm{s,f}=\frac{\sqrt{2\gamma(\gamma-1)}(\alpha-1)}{(\gamma+1)}At^{\alpha},\ \ \ 
\rho_\mathrm{f}=\frac{\gamma+1}{\gamma-1}A^{-m}Dt^{\delta}.
\label{eq:jump_condition}
\end{equation}
\subsection{Governing Equations}
The equations of hydrodynamics for non-relativistic spherically symmetric flows are described as follows,
\begin{equation}
\frac{\partial\rho}{\partial t}+\frac{\partial(\rho v)}{\partial R}+\frac{2\rho v}{R}=0,
\label{eq:cont_spherical}
\end{equation}
\begin{equation}
\frac{\partial u}{\partial t}+u\frac{\partial u}{\partial R}+\frac{1}{\gamma\rho}\frac{\partial (\rho c_\mathrm{s}^2)}{\partial R}=0,
\end{equation}
and
\begin{equation}
\frac{\partial (\rho^{1-\gamma}c_\mathrm{s}^2)}{\partial t}
+u\frac{\partial (\rho^{1-\gamma}c_\mathrm{s}^2)}{\partial R}
=0.
\end{equation}
We define the similarity variable $\xi$ as the ratio of the radial coordinate to the shock radius,
\begin{equation}
\xi=\frac{R}{R_\mathrm{s}}=\frac{R}{At^\alpha},
\end{equation}
and the dimensionless velocity $U$, sound speed $C$, and density $G$, which are functions of the similarity variable, as follows,
\begin{equation}
u=\alpha At^{\alpha-1}\xi U(\xi)
,\ \ \ 
c_\mathrm{s}=\alpha At^{\alpha-1}\xi C(\xi)
,\ \ \ 
\rho=A^{-m}Dt^{\delta}G(\xi).
\end{equation}
From the transformation of the independent variables from $(t,r)$ to $(t,\xi)$, the operators $\partial/\partial t$ and $\partial/\partial R$ appearing in the equations of hydrodynamics are transformed into the following forms,
\begin{equation}
\frac{\partial }{\partial t} \rightarrow\frac{\partial}{\partial t}-\alpha\frac{\xi}{t}\frac{\partial}{\partial \xi},\ \ \ 
\frac{\partial }{\partial R}\rightarrow\frac{1}{At^\alpha}\frac{\partial }{\partial \xi}.
\end{equation}
Using these relations, the equations of hydrodynamics can be transformed into the following ordinary differential equations,
\begin{equation}
\alpha(U-1)\frac{d\ln G}{d\ln\xi}+\alpha\frac{dU}{d \ln\xi}+3\alpha U+\delta=0,
\label{eq:mass}
\end{equation}
\begin{equation}
\alpha(U-1)\frac{dU}{d\ln \xi}+(\alpha U-1)U+\frac{\alpha C^2}{\gamma}\left[2+\frac{d\ln G}{d\ln \xi}+2\frac{d\ln C}{d\ln\xi}\right]=0,
\label{eq:momentum}
\end{equation}
and
\begin{equation}
\alpha(1-\gamma)(U-1)\frac{d\ln G}{d\ln \xi}+2\alpha(U-1)\frac{d\ln C}{d\ln \xi}+
2(\alpha U-1)+(1-\gamma)\delta=0.
\label{eq:energy}
\end{equation}

We further transform these equations before solving them. 
Equations (\ref{eq:mass}) and (\ref{eq:energy}) are combined to give the following equation,
\begin{equation}
[2(1-\alpha)+(\gamma-1)\delta]\frac{dU}{d\ln\xi}+
2(3\alpha+\delta)(U-1)\frac{d\ln C}{d\ln\xi}+
(2+\alpha-3\alpha\gamma)\frac{d\ln G}{d\ln\xi}+
(6-\delta+3\delta\gamma)(U-1)=0,
\end{equation}
which can be integrated to give the following relation between the dimensionless functions $U$, $C$, and $G$,
\begin{equation}
|1-U|^{2(1-\alpha)+(\gamma-1)\delta}C^{2(3\alpha+\delta)}G^{2+\alpha-3\alpha\gamma}
\xi^{6-\delta+3\delta\gamma}=\mathrm{Const.}
\label{eq:G}
\end{equation}
Therefore, the function $G$ is obtained for a given set of $U$ and $C$ at $\xi$. 
Next, the elimination of $d\ln G/d\ln\xi$ in Equations (\ref{eq:momentum}) and (\ref{eq:energy}) by using Equation (\ref{eq:mass}) with some algebraic manipulations yields
\begin{equation}
\frac{dU}{d\ln\xi}=\frac{\Delta_1}{\Delta},\ \ \ 
\frac{d\ln C}{d\ln\xi}=\frac{\Delta_2}{\Delta},
\label{eq:govern}
\end{equation}
where
\begin{equation}
\Delta=(1-U)^2-C^2,
\end{equation}
\begin{equation}
\Delta_1=U(1-U)\left(U-\frac{1}{\alpha}\right)+\left(3U-\frac{2-2\alpha-\delta}{\alpha\gamma}\right)C^2,
\label{eq:delta_1}
\end{equation}
and
\begin{equation}
\Delta_2=
(U-1)\left(\frac{3\gamma-1}{2}U-\frac{1}{\alpha}\right)+\frac{\gamma-1}{2}U\left(U-\frac{1}{\alpha}\right)+C^2+
\frac{2(1-\alpha)+\delta(\gamma-1)}{2\alpha\gamma}\frac{C^2}{1-U}.
\end{equation}
\subsection{Boundary Conditions}
The values of the dimensionless functions $U_\mathrm{f}$, $C_\mathrm{f}$, and $G_\mathrm{f}$ at the shock front $\xi=1$ are found from the shock jump conditions (\ref{eq:jump_condition}),
\begin{equation}
U_\mathrm{f}=\frac{2\alpha+\gamma-1}{(\gamma+1)\alpha},\ \ \ 
C_\mathrm{f}=\frac{\sqrt{2\gamma(\gamma-1)}(\alpha-1)}{(\gamma+1)\alpha},\ \ \ 
G_\mathrm{f}=\frac{\gamma+1}{\gamma-1}.
\end{equation}
We use these values as the initial values when we integrate the governing equations from $\xi=1$. 

Next, we consider the inner boundary conditions. 
Equation (\ref{eq:govern}) diverges when the denominator $\Delta $ vanishes, i.e.,
\begin{equation}
U+C=1.
\label{eq:sonic_condition}
\end{equation}
This condition divides the $U$-$C$ plane into subsonic ($U+C>1$) and supersonic ($U+C<$1) regions. 
In the subsonic region, the characteristic velocity is higher than the shock velocity and thus information at a point in the post-shock region can catch up with the shock front in a finite time. 
On the other hand, from any point in the supersonic region, no information can affect the shock front. 

The numerator $\Delta_1$ in the first of Equation (\ref{eq:govern}) should be zero at the same time when the condition (\ref{eq:sonic_condition}) is satisfied so that the differential equations can be integrated successfully. 
We define the variables $(U_\mathrm{cr},C_\mathrm{cr})$ satisfying $\Delta=0$ and $\Delta_1=0$ simultaneously. 
Substituting $C_\mathrm{cr}=(1-U_\mathrm{cr})$ into $\Delta_1=0$, one obtains
\begin{equation}
U_\mathrm{cr}(1-U_\mathrm{cr}) \left(U_\mathrm{cr}-\frac{1}{\alpha}\right)
+(1-U_\mathrm{cr})^2\left(3U_\mathrm{cr}-\frac{2-2\alpha-\delta}{\alpha\gamma}\right)=0.
\label{eq:sonic_point}
\end{equation}
\subsection{Subsonic Solution}
The following set satisfies Equation (\ref{eq:sonic_point}),
\begin{equation}
(U_\mathrm{cr},C_\mathrm{cr})=(1,0).
\end{equation}
To find the solution corresponding to this set, we integrate the following equation from the shock front $(U_\mathrm{f},C_\mathrm{f})$ to the sonic point $(U_\mathrm{cr},C_\mathrm{cr})$,
\begin{equation}
\frac{d\ln(1-U)}{d\ln C}=-\frac{\Delta_1}{(1-U)\Delta_2}.
\label{eq:dUdlnC}
\end{equation} 
The integration can be performed in a straightforward way, because no singular point is located on the path of the integration. 
At the same time, the reciprocal of the second of Equation (\ref{eq:govern}) is integrated from $(\xi,C)=(1,C_\mathrm{f})$ to find the value $\xi_\mathrm{c}$ of the similarity variable at the sonic point. 
Generally, $\xi_\mathrm{c}$ takes a non-zero value in the range $0<\xi_\mathrm{c}<1$, which means that the solution is truncated at a certain point, $R_\mathrm{c}=\xi_\mathrm{c}R_\mathrm{f}$, between the origin and the shock front. 
This radius is interpreted as the interface between the shocked ejecta and the shocked wind, i.e., the contact discontinuity, where we impose a boundary condition. 

Before introducing the boundary condition, we consider the asymptotic behaviour of the solution around $\xi=\xi_\mathrm{c}$. 
When $\xi$ is close to $\xi_\mathrm{c}$, $(1-U)$ and $C$ are much smaller than unity. 
Thus, we approximate Equation (\ref{eq:dUdlnC}) under $1-U,C\ll 1$. 
Assuming that $C^2$ approaches to zero more slowly than $(1-U)$ when $\xi\rightarrow \xi_\mathrm{c}$, the leading terms in $\Delta_1$ and  $\Delta_2$ are proportional to $C^2$ and $C^2/(1-U)$, respectively.  
Then, Equation (\ref{eq:dUdlnC}) can be approximated as follows,
\begin{equation}
\frac{d\ln(1-U)}{d \ln C}\simeq \frac{2[2(1-\alpha)-\delta-3\alpha\gamma]}{2(1-\alpha)+\delta(\gamma-1)}.
\end{equation}
This expression can be integrated to give the following relation,
\begin{equation}
(1-U)\propto C^{2\beta},
\end{equation}
where
\begin{equation}
\beta=1-\frac{\gamma(3\alpha+\delta)}{2(1-\alpha)+\delta(\gamma-1)}.
\end{equation}
The exponent $\beta$ is found to be larger than unity for parameters of interest, which justifies the assumption that $C^2$ approaches to zero more slowly than $(1-U)$ when $\xi\rightarrow\xi_\mathrm{c}$. 
Instituting this relation into Equation (\ref{eq:G}), one finds that the following relation,
\begin{equation}
(C^2G)^{2+\alpha-3\alpha\gamma}\xi^{6-\delta+3\delta\gamma}=\mathrm{Const.},
\end{equation}
holds around $\xi=\xi_\mathrm{c}$. 
Since $C^2$ vanishes at $\xi=\xi_\mathrm{c}$, the dimensionless density $G$ diverges to infinity. 
However, the above relation shows that the product $C^2G$, which is proportional to the pressure $p=\gamma c_\mathrm{s}^2\rho$, approaches to a non-zero value at $\xi=\xi_\mathrm{c}$, which guarantees the finiteness of the pressure at the contact discontinuity. 
We define the dimensionless pressure $\eta_\mathrm{c}$ at the contact discontinuity as the following limit,
\begin{equation}
\eta_\mathrm{c}=\lim_{\xi\rightarrow\xi_\mathrm{c}}\xi^2C(\xi)^2G(\xi).
\end{equation}
Using this quantity, which is found from the numerical integration of the governing equations, the pressure at the contact discontinuity is expressed as follows,
\begin{equation}
p_\mathrm{c}=\alpha^2\gamma A^{2-m}D\eta_\mathrm{c}t^{2\alpha+\delta-2}.
\label{eq:pc_ejecta}
\end{equation}

The contact discontinuity connects the shocked ejecta and the shocked relativistic wind. 
Equating the pressure at the bottom of the shocked ejecta, Equation (\ref{eq:pc_ejecta}), and that of the hot bubble, Equation (\ref{eq:pb}), one finds the expressions of the exponent $\alpha$ and the normalisation constant $A$,
\begin{equation}
\alpha=\frac{6-m}{5-m},
\end{equation}
and
\begin{equation}
A^{5-m}=\frac{3(\gamma-1)(2-\gamma)L}{\alpha^2\gamma[1+3\alpha(\gamma-1)] \xi_\mathrm{c}^3\eta_\mathrm{c}(\omega_\mathrm{c}v_\mathrm{ej})^{m-3}f_\mathrm{3}M_\mathrm{ej}}.
\end{equation}

\section{SPECIAL RELATIVISTIC HYDRODYNAMICS CODE WITH ADAPTIVE MESH REFINEMENT}\label{sec:code}
In our simulations, we use a special relativistic hydrodynamics code developed by one of the authors. 
In this section, we briefly describe the numerical code. 

The code is an extension of the one-dimensional version presented in the previous paper \citep{2014ApJ...796...30S}. 
We adopt the 3rd-order MUSCL scheme \citep{1977JCoPh..23..276V,1979JCoPh..32..101V} for the spatial reconstruction and the HLLC scheme to evaluate the numerical fluxes at cell surfaces \citep{2005MNRAS.364..126M}. 
We use the HLLC scheme presented by \cite{toro2013} for non-relativistic problems. 
For the time integration, we adopt the 3rd-order TVD Runge-Kutta method \citep{1988JCoPh..77..439S}.

Our AMR code is based on the so-called block-structured technique, in which a number of AMR blocks composed of $8\times8$ two-dimensional numerical cells cover the whole numerical domain. 
An AMR block (parent block) is divided into 4 child blocks if the parent block needs higher resolution. 
Our refinement criteria are similar to those employed by \cite{2000ApJS..131..273F}. 
For a physical variable $q$, the following quantity is evaluated for a cell along each direction by using $q_i$ at the cell and those, $q_{i+1}$ and $q_{i-1}$, at the neighbouring cells,
\begin{equation}
E_\mathrm{amr}=\frac{|q_{i+1}-2q_i+q_{i-1}|}{|q_{i+1}-q_{i}|+|q_{i}-q_{i-1}|+\epsilon_\mathrm{amr}|q_{i+1}-2q_i+q_{i-1}|},
\end{equation}
with $\epsilon_\mathrm{amr}=0.01$. 
We obtain the maximum value of this quantity among all the numerical cells in an AMR block and determine whether the AMR block needs to be covered by child blocks or not, depending on the value. 
Practically, we employ a threshold $E_\mathrm{max}=0.8$, above which finer resolution is required. 
On the other hand, when the values of all the child blocks covering a parent block are smaller than another threshold $E_\mathrm{min}=0.2$, the child blocks are integrated to the parent block. 
We usually use the density and pressure to calculate the quantity $E_\mathrm{amr}$. 

In the previous paper \citep{2014ApJ...796...30S}, we demonstrate that the one-dimensional version of our code can successfully solve some test problems, the Sod's shock tube problem, the Einfeldt's strong rarefaction test, the Sedov-Taylor point explosion test, and the relativistic shock tube problem. 
Thus, we show results of two-dimensional test problems in the following.

\subsection{Double Mach Reflection}
This test, which was first proposed by \cite{1984JCoPh..54..115W}, deals with a strong shock wave reflecting at a wall. 
The simulation is carried out in a two-dimensional cartesian space with $0\leq x\leq 4$ and $0\leq y\leq 1$. 
At the lowest resolution, the numerical domain is covered by $128\times32$ numerical cells. 
The maximum AMR level is set to $l_\mathrm{max}=4$, which means that the numerical domain is effectively covered by $2048\times 512$ numerical cells. 
The adiabatic index of the gas is set to $1.4$ in this test problem. 
Initially, a boundary at $x=x_0+y\tan(\pi/6)$ with $x_0=1/6$ separates a shocked medium from an unshocked medium at rest. 
The shock is propagating in a direction with an angle of $30^\circ$ from the $x$-axis. 
Thus, the initial conditions for the physical variables are as follows,
\begin{equation}
(\rho,v_x,v_y,p)=
\left\{
\begin{array}{cl}
(8.0,8.25\cos(\pi/6),-8.25\sin(\pi/6),116.5)&\mathrm{for}\ x<x_0+y\tan(\pi/6),\\
(1.4,0,0,1)&\mathrm{for}\ x\geq x_0+y\tan(\pi/6).
\end{array}\right.
\end{equation}
For the boundary at $y=0$, the following condition is imposed to make the shock reflected,
\begin{equation}
(\rho,v_x,v_y,p)=
\left\{
\begin{array}{cl}
(8.0,8.25\cos(\pi/6),-8.25\sin(\pi/6),116.5)&\mathrm{for}\ x<x_0+(1+20t)\tan(\pi/6),\\
(1.4,0,0,1)&\mathrm{for}\ x\geq x_0+(1+20t)\tan(\pi/6),
\end{array}\right.
\end{equation}
where $t$ is the elapsed time. 
For the other boundaries, free boundary conditions are imposed. 

\begin{figure*}[tbp]
\begin{center}
\includegraphics[scale=0.5,bb=0 0 709 510]{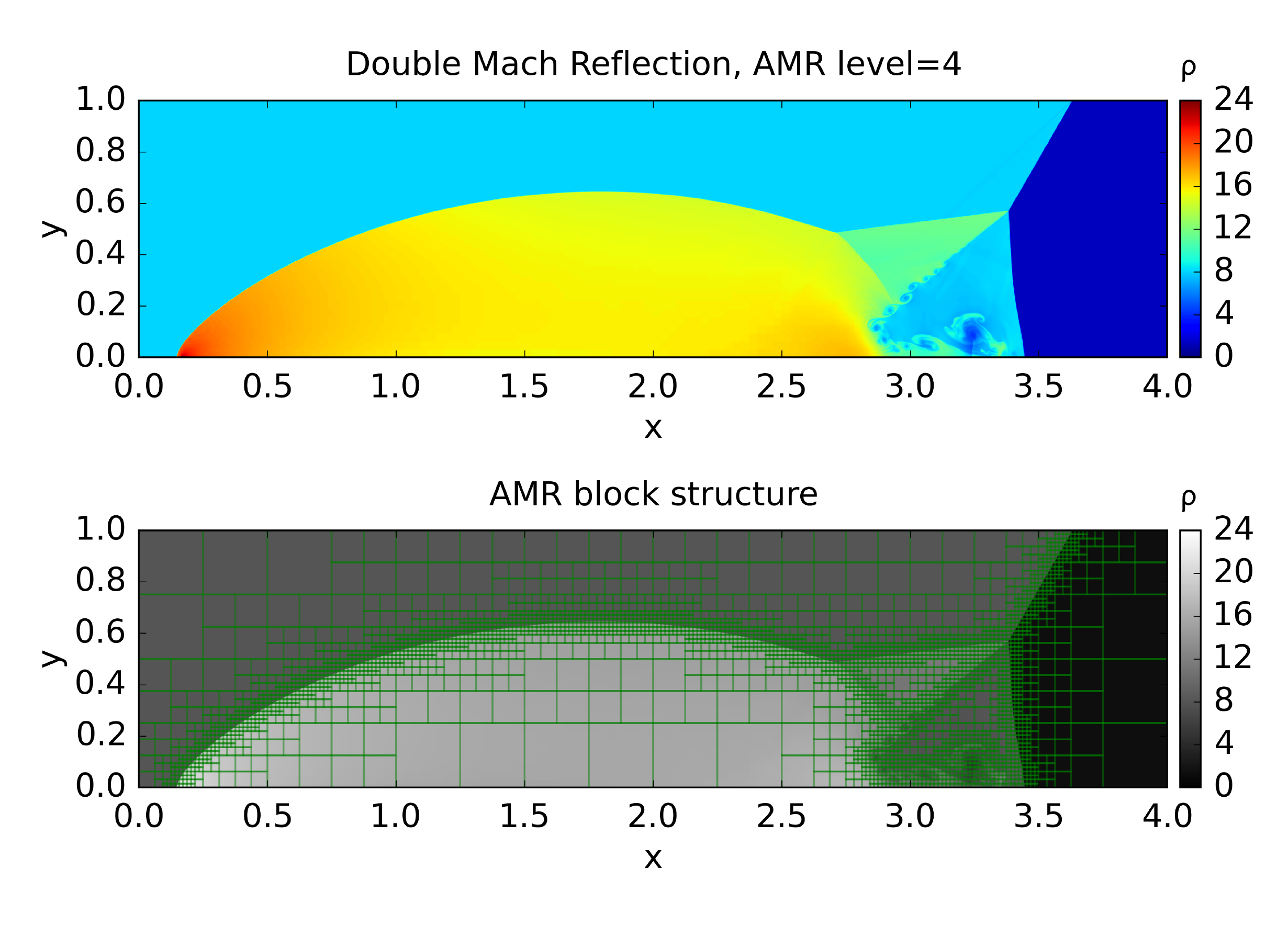}
\caption{Results of the double mach reflection test. 
In the upper panel, the colour-coded density distribution at $t=0.25$ is plotted. 
In the lower panel, the same distribution is plotted in grayscale and green boxes in the the plot represent AMR blocks covering the numerical domain.}
\label{fig:double_mach}
\end{center}
\end{figure*}

Figure \ref{fig:double_mach} shows the results of the test problem. 
The upper panel shows the spatial distribution of the density at $t=0.25$, while the lower panel shows the same distribution in grayscale and the structure of AMR blocks covering the domain at the time. 
After the simulation starts, the shock wave with a Mach number of $10$ propagates in a static gas and then reaches the reflecting wall. 
As a result, a two-dimensional self-similar flow is created. 
The result of this test problem is widely known and our simulation well reproduces the result. 
Especially, our numerical code efficiently follows the evolution of the discontinuities developing in the flow. 
As seen in the lower panel of Figure \ref{fig:double_mach}, which shows the grid structure realized in the simulation, the shocked and unshocked regions without complicated structure are covered by AMR blocks with the lowest resolution, while shock fronts are covered by AMR blocks with the highest resolution.

\subsection{2D Special Relativistic Shock Tube}
\begin{figure*}[tbp]
\begin{center}
\includegraphics[scale=0.6,bb=0 0 850 397]{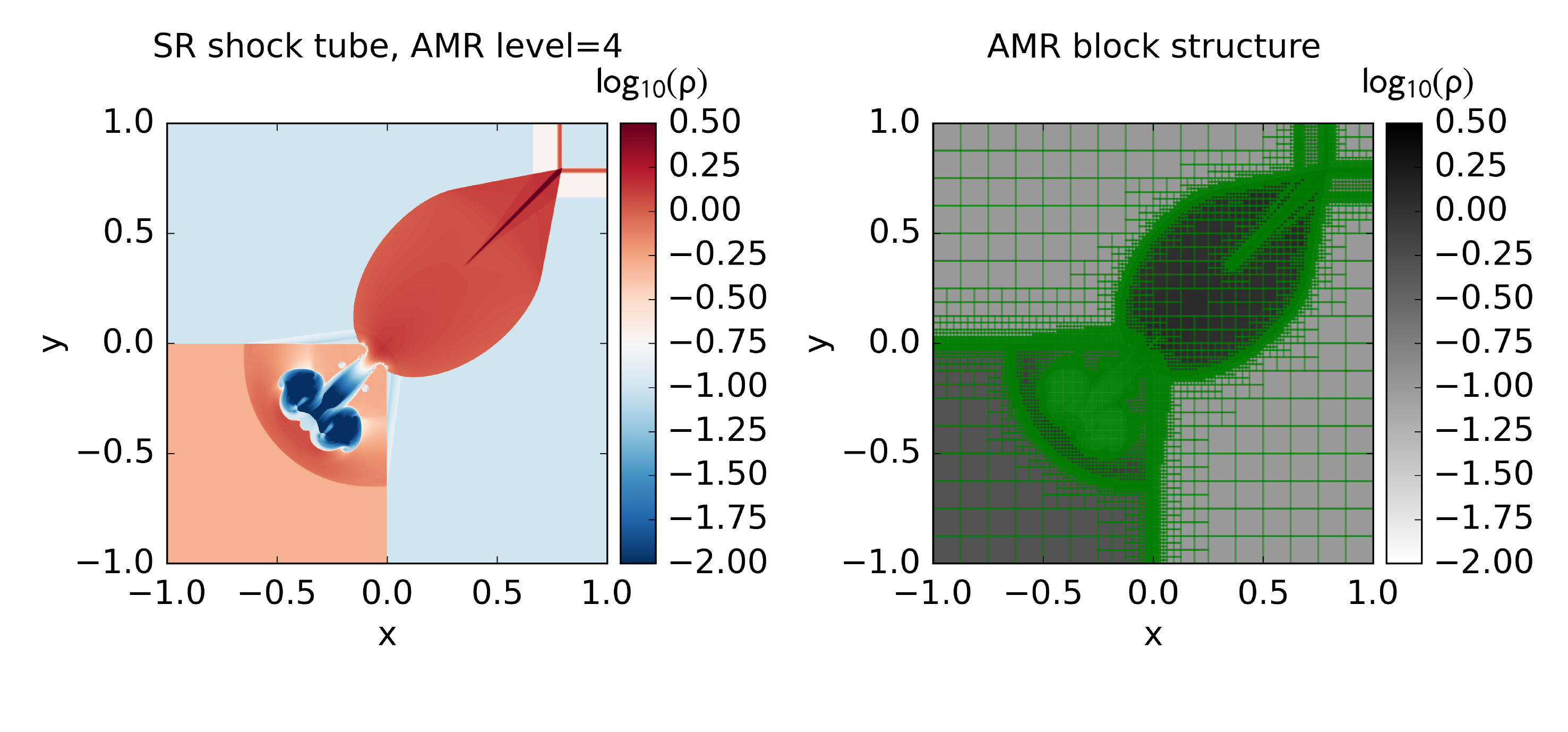}
\caption{Results of the 2D special relativistic shock tube test. 
In the left panel, the color-coded density distribution at $t=0.8$ is plotted. 
In the right panel, the same distribution is plotted in grayscale and green boxes in the the plot represent AMR blocks covering the numerical domain.}
\label{fig:sr_shocktube}
\end{center}
\end{figure*}

In this test, we consider the formation and propagation of relativistic shock waves in two-dimensional cartesian space with $-1\leq x\leq 1$ and $-1\leq y\leq 1$. 
The numerical domain is covered by $128\times 128$ cells at the lowest AMR level $l=0$. 
The maximum refinement level is set to $l_\mathrm{max}=4$. 
Thus, the domain is effectively covered by $2048\times2048$ numerical cells. 
The adiabatic index of the gas is set to $5/3$. 
We impose the following initial conditions for the primitive variables,
\begin{equation}
(\rho,v_x,v_y,p)=
\left\{
\begin{array}{ccccc}
(0.1,0,0,0.01)&\mathrm{for}&x<0&\mathrm{and}&y<0,\\
(0.1,0.99,0,0.1)&\mathrm{for}&x<0&\mathrm{and}&0\leq y,\\
(0.1,0,0.99,0.1)&\mathrm{for}&0\leq x&\mathrm{and}&y<0,\\
(0.5,0,0,0.1)&\mathrm{for}&0\leq x&\mathrm{and}&0\leq y,
\end{array}\right.
\end{equation}
and free boundary conditions for the boundaries in both $x$ and $y$. 
After the simulation starts, the relativistic flows in $x$ and $y$ directions create shock waves and they propagate into the flows. 
The same test problem is solved by several authors \citep[e.g.][]{2002A&A...390.1177D,2004A&A...428..703L,2005MNRAS.364..126M}. 
Results obtained by our codes is shown in Figure \ref{fig:sr_shocktube}, in which the colour-coded density distribution and the corresponding grid structure are presented. 
We confirm that the results well agrees with earlier studies. 
We can also see in the right panel of Figure \ref{fig:sr_shocktube} that the shock fronts are efficiently covered by AMR blocks with higher resolution thanks to the AMR technique. 

\subsection{Relativistic Jet Propagation}
\begin{figure*}[tbp]
\begin{center}
\includegraphics[scale=0.5,bb=0 0 1020 453]{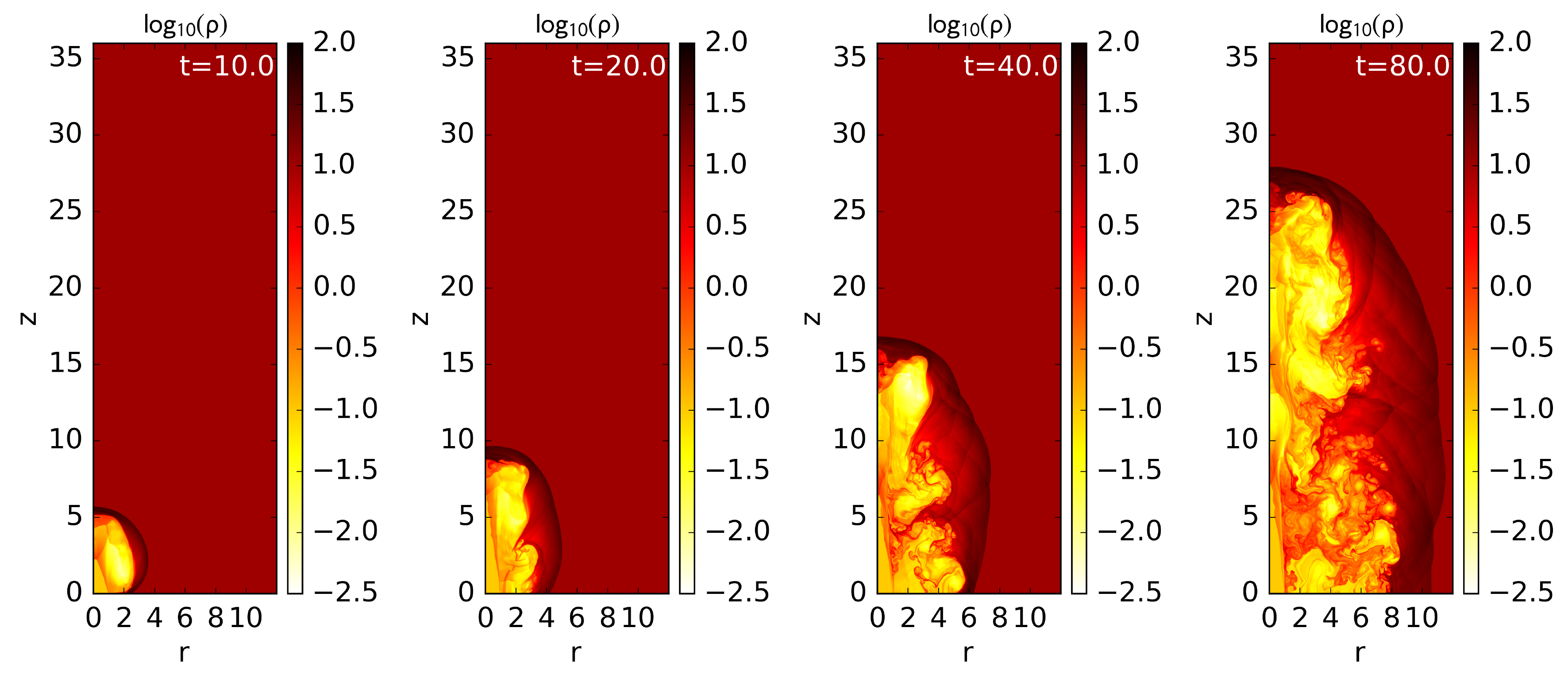}
\caption{Results of the relativistic jet propagation test. 
The panels show the color-coded spatial distributions of the density at $t=10$, $20$, $40$, and $80$ from left to right. }
\label{fig:jet}
\end{center}
\end{figure*}

This test is carried out with two-dimensional cylindrical coordinates $(r,z)$, which cover  a region with $0\leq r\leq 12$ and $0\leq z\leq 36$. 
The numerical domain is covered by $128\times384$ numerical cells at the lowest level and the maximum refinement level is $l_\mathrm{max}=4$. 
The adiabatic index of the gas is set to $5/3$. 
A jet travelling at a relativistic speed $v_z=0.99$ along the $z$-axis is continuously injected at the bottom. 
The initial conditions for the primitive variables are as follows,
\begin{equation}
(\rho,v_r,v_z,p)=
\left\{
\begin{array}{cl}
(0.1,0,0.99,0.01)&\mathrm{for}\ r<1\ \mathrm{and}\ z<1,\\
(1.0,0,0,0.01)&\mathrm{otherwise}.
\end{array}\right.
\end{equation}
The boundary condition at $r<1$ and $z=0$ is
\begin{equation}
(\rho,v_r,v_z,p)=
(0.1,0,0.99,0.01),
\end{equation}
so that the jet injection is maintained. 
A reflecting boundary condition is imposed at the symmetry axis $z=0$, while free boundary conditions are imposed for the other boundaries. 
Figure \ref{fig:jet} shows the evolution of the jet. 
The panels in Figure \ref{fig:jet} show spatial distributions of the density at several epochs. 
The jet propagates in a uniform ambient medium with a density higher than the jet, creating a strong blast wave. 
In the shocked region, the Kelvin-Helmholtz instability develops at the interface between the jet and the shocked ambient medium. 
As a result, the shocked gas is subject to efficient mixing.  
The minimum scale of the structure created by the instability is determined by the resolution of the simulation. 
Therefore, small structures created inside the jet are different from earlier studies dealing with similar problems \citep{2002A&A...390.1177D,2004A&A...428..703L,2005MNRAS.364..126M}. 
Nevertheless, the overall picture of the jet propagation reproduced by our simulation well agrees with theirs.

\end{document}